\documentclass[submission, Phys]{SciPost}
\pdfoutput=1
\usepackage{comment}
\usepackage{enumerate}
\usepackage{amssymb}
\usepackage{amsmath}
\usepackage{braket}
\usepackage{graphicx}

\newcommand{\be}{\begin{equation}}
\newcommand{\ee}{\end{equation}}
\newcommand{\ba}{\begin{aligned}}
\newcommand{\ea}{\end{aligned}}
\newcommand{\bw}{\begin{widetext}}
\newcommand{\ew}{\end{widetext}}

\renewcommand{\vec}[1]{\boldsymbol{#1}}
\newcommand{\bea}{\begin{eqnarray}}
\newcommand{\eea}{\end{eqnarray}}

\begin{document}

\begin{center}{\Large \textbf{
Unbounded entanglement production via a dissipative impurity
}}\end{center}

% TODO: write the author list here. Use initials + surname format.
% Separate subsequent authors by a comma, omit comma at the end of the list.
% Mark the corresponding author with a superscript *.
\begin{center}
Vincenzo Alba\textsuperscript{1*}
\end{center}

\begin{center}
{\bf 1}	Institute  for  Theoretical  Physics, Universiteit van Amsterdam,
Science Park 904, Postbus 94485, 1098 XH Amsterdam,  The  Netherlands
\\
* v.alba@uva.nl
\end{center}

\begin{center}
\today
\end{center}

% For convenience during refereeing: line numbers
%\linenumbers

\section*{Abstract}
{\bf
We investigate the entanglement dynamics in a free-fermion chain initially 
prepared in a Fermi sea and subjected to 
localized losses (dissipative impurity). We derive a formula describing the dynamics 
of the entanglement entropies  in 
the hydrodynamic limit of long times and large intervals. 
The result depends only on the absorption coefficient of the effective 
delta potential describing the impurity in the hydrodynamic limit. 
Genuine dissipation-induced entanglement  is certified by  
the linear growth of the logarithmic negativity. 
Finally, in the quantum Zeno regime  at strong dissipation the entanglement growth is 
arrested (Zeno entanglement death). 
}

%#############################################################
\section{Introduction}
\label{sec:intro}

Common experience suggests that  the interaction between a quantum system 
and its environment, and the ensuing dissipation,  is 
detrimental for quantum entanglement. In recent 
years this view was challenged as it was realized that 
dissipation can be a resource to engineer quantum states~\cite{lin2013}, 
for quantum computation~\cite{verstraete-2009}, or to stabilize exotic 
states of matter, such as topological order~\cite{diehl-2011}. 
These results, together with  the interest in Noisy-Intermediate-Scale-Quantum 
(NISQ) devices\cite{preskill2018quantum}, urge for a thorough understanding of the interplay 
between entanglement and dissipation in open quantum systems. 

A major obstacle is that it is a challenging task to encapsulate the 
system-environment interaction  within a theoretical framework. 
Within the so-called Markovian approximation,  the Lindblad equation provides a powerful 
framework to address open quantum systems~\cite{petruccione}. 
Interestingly, for some models it is possible to obtain exact solutions 
of the Lindblad equation~\cite{prosen-2008,prosen-2011,prosen-2014,prosen-2015,
znidaric-2010,znidaric-2011,medvedyeva-2016,buca-2020,bastianello-2020,essler-2020,ziolkowska-2020},  
for instance, in noninteracting systems with {\it linear} dissipators~\cite{prosen-2008}. 
Perturbative field-theoretical approaches are also available~\cite{sieberer-2016}. 
A promising direction is to extend the hydrodynamic framework to integrable systems 
subjected to dissipation~\cite{bouchoule-2020,bastianello-2020,Friedman_2020,deleeuw-2021,denardis2021}. 
This is motivated by the tremendous success of Generalized Hydrodynamics (GHD)  
for integrable systems~\cite{bertini-2016,olalla-2016}.  
In some simple free-fermion setups it has been shown that it is possible to 
use a hydrodynamic approach to described the entanglement dynamics~\cite{alba-2021,maity-2020}.  
This generalizes a well-known quasiparticle picture for the entanglement 
spreading in integrable systems~\cite{calabrese-2005,fagotti-2008,alba-2017,alba-2018,alba2017quench,alba2017renyi,alba2018entanglement,alba2021generalizedhydrodynamic}.  

%
%########################################
\begin{figure}[t]
	\begin{center}
\includegraphics[width=1\textwidth]{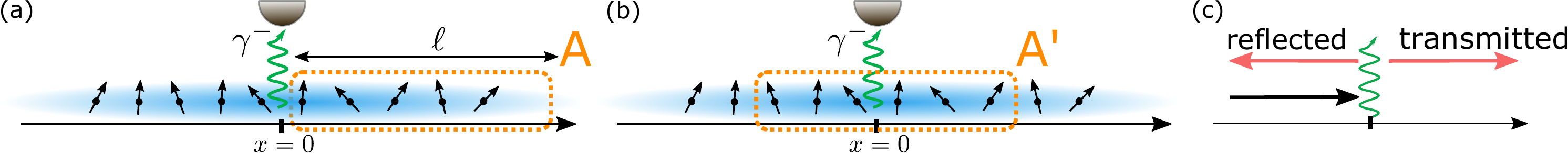}
\caption{ Dissipation-induced entanglement growth. (a) and (b) A free-fermion chain 
 is prepared in Fermi sea and subject to fermionic losses acting at the 
 center of the chain at $x=0$. Here $\gamma^-$ is the loss rate. 
 We are interested in the entanglement entropy $S$ of a subregion  of 
 length $\ell$. We consider two partitions. In the first one (side partition) 
 $A$ is placed  next to the dissipative impurity (see (a)). In the second one (centered 
 partition) a subregion $A'$  is 
 centered around the impurity (see (b)). (c) Mechanism 
 for entanglement generation. A fermion reaching the origin can be 
 absorbed or reflected. The reflected and transmitted fermions are entangled.  
}
\end{center}
\label{fig0:cartoon}
\end{figure}
%########################################
%

Dissipative impurities provide a minimal theoretical laboratory to study the 
effects of dissipation  in quantum many-body systems. 
They are the focus of rapidly-growing interest, both theoretical~\cite{dolgirev-2020,jin-2020,maimbourg-2020,
froml-2019,tonielli-2019,froml-2020,krapivsky-2019,krapivsky-2020,rosso-2020,vernier-2020,alba2021noninteracting,chaudhari2021zeno,muller-2021}, as well as 
experimental~\cite{gericke-2008,brazhnyi-2009,zezyulin-2012,barontini-2013,patil-2015,labouvie-2016}.   
The interplay between entanglement and thermodynamic entropy in the presence of dissipative 
impurities has not been explored much. 

One aim of this paper is to start such investigation. We focus on 
noninteracting fermions with localized fermion losses. The chain is initially prepared 
in a Fermi sea, and then undergoes Lindblad dynamics. To monitor the entanglement dynamics 
we consider the 
entanglement entropies~\cite{amico2008entanglement,calabrese2009entanglemententropy,eisert2010colloquium,laflorencie2016quantum} 
(both von Neumann and R\'enyi entropies), and the fermionic 
logarithmic negativity~\cite{eisert1999a,lee2000partial,vidal2002computable,plenio2005logarithmic,ruggiero2016negativity,ruggiero2016entanglement,wichterich2009scaling,marcovitch2009critical,calabrese2012entanglement,calabrese2013entanglement,
coser2014entanglement,eisler2014entanglement,eisler2015partial,blondeau2016universal,shapourian2017many,shapourian2017partial,shapourian2019entanglement}. 
The setup is depicted in Fig.~\ref{fig0:cartoon}. An infinite chain is prepared in a Fermi 
sea with generic Fermi level $k_F$. The dissipation acts at the origin removing fermions 
incoherently at a rate $\gamma^-$. To quantify the entanglement shared between different subregions we consider 
the bipartitions of the chain shown in Fig.~\ref{fig0:cartoon} (a) and (b). 
In (a) (side bipartition) a subsystem $A$ of length $\ell$ is placed next to the impurity, whereas 
in (b) (centered partition) a subsystem $A'$ of the same length is centered 
around the origin. 
Here we focus on the hydrodynamic limit of large $\ell$ and long times, with their ratio fixed. 
A crucial observation is that in the hydrodynamic limit of large distances 
from the dissipation source and long times, the dissipation acts as an effective delta 
potential (dissipative impurity) with imaginary strength. The 
associated reflection and transmission amplitudes can be derived 
analytically~\cite{alba2021noninteracting}. The presence of loss dissipation 
is reflected in a nonzero absorption coefficient.

Due to the nonunitary dynamics  entanglement and thermodynamic correlations 
are deeply intertwined. The origin of entanglement is understood as follows. The mechanism 
is depicted in Fig.~\ref{fig0:cartoon} (c). The effective delta potential at the origin 
gives rise to a superposition between the transmitted and the reflected fermion, which form 
an entangled pair. The propagation of entangled pairs generate 
entanglement between different spatial regions of the system. More precisely, regions that 
share entangled pairs get entangled.  A similar mechanism is responsible for entanglement 
production in free-fermion chains with a defect~\cite{eisler2012entanglement,collura2013entanglement,gruber2020time,gamayun-2020,gamayun2021nonequilibrium}. 
Together with quantum entanglement, thermodynamic correlation is produced during the dynamics. 
Although the initial state is homogeneous, dissipation gives rise to a nontrivial 
density profile. This is accompanied by the creation of thermodynamic entropy. Here we show that the 
entanglement entropies cannot distinguish between these two types of correlation. Indeed, quite generically 
the entanglement entropies grow linearly with time. This linear growth of the von Neumann entropy 
in open quantum systems has been observed already, for instance, in~\cite{ptaszy2019entropy}. 
One of our main results is that in the hydrodynamic 
limit the entanglement entropy of $A$ (see Fig.~\ref{fig0:cartoon} (a)) is described by 
\begin{equation}
	\label{eq:ent-hydro-intro}
	S=\frac{\ell}{2}\int_{-k_F}^{k_F}\frac{dk}{2\pi} H_1(1-|a|^2)\min(|v_k|t/\ell,1). 
\end{equation}
We provide similar results for $A'$. 
In~\eqref{eq:ent-hydro-intro} we defined  
$H_1(x):=-x\ln(x)-(1-x)\ln(1-x)$, and $v_k$ is the fermion group 
velocity. Crucially, $|a|^2$ is the absorption coefficient, which is nonzero because of the losses. 
For lattice systems a maximum velocity $v_\mathrm{max}$ exists and~\eqref{eq:ent-hydro-intro} predicts 
a linear  growth at short times $v_\mathrm{max}t/\ell<1$, followed by a volume-law 
scaling at long times. We provide similar results for the R\'enyi entropies and the moments 
of fermionic correlation functions. 
Formula~\eqref{eq:ent-hydro-intro} is similar to that describing the entanglement 
dynamics in a free-fermion chain with a bond defect~\cite{eisler2012entanglement}. The main difference 
is that in the 
unitary case the growth of the entropy depends only on the 
transmission coefficient of the defect. We should stress that although we present results only 
for the two geometries in Fig.~\ref{fig0:cartoon}, it should be possible to generalize Eq.~\eqref{eq:ent-hydro-intro} 
to to arbitrary bipartitions. 

Again, the linear growth in~\eqref{eq:ent-hydro-intro} does not reflect genuine entanglement production, 
which can be diagnosed by the logarithmic negativity. 
For instance, we show that the logarithmic negativity grows linearly with time for subsystem $A$, whereas 
it does not increase for $A'$. This supports the mechanism outlined above. For the bipartition in 
Fig.~\ref{fig0:cartoon} entanglement is due to the shared pairs formed by the transmitted and the 
reflected fermions. On the other hand, for the bipartition in Fig.~\ref{fig0:cartoon} (b) these 
pairs are never shared between $A'$ and its complement. 

The manuscript is organized as follows. In section~\ref{sec:intro} we introduce 
the model and the setup. In particular, we review the formula for the fermionic correlators 
in the hydrodynamic limit, which are the main ingredients to compute the entanglement entropies and 
the negativity. Entangled-related quantities are introduced in section~\ref{sec:obs}. In section~\ref{sec:hydro-ent} we 
present our main results. We first discuss the formula describing 
arbitrary functions of the moments of the fermionic correlators in the hydrodynamic limit. 
In section~\ref{sec:mn} we specialize to the moments of the fermionic correlators. 
In section~\ref{sec:ent-theo} we discuss the hydrodynamic behavior of the entanglement entropies. 
In section~\ref{sec:ss-entropy} we focus on the stationary value of the 
entanglement entropy, discussing its dependence on  the dissipation strength. 
In section~\ref{sec:num} we present numerical benchmarks. We focus on the moments of the 
fermionic correlators in section~\ref{sec:num-a}, and on the entanglement entropies in 
section~\ref{sec:num-b}. We discuss some  future directions in section~\ref{sec:concl}. 
In Appendix~\ref{sec:app} we report the derivation of the main result of section~\ref{sec:hydro-ent}.

%#############################################################
\section{Localized losses in a Fermi sea}
\label{sec:intro}

Here we consider the infinite free-fermion chain defined by the Hamiltonian 
\begin{equation}
\label{eq:ham}
	H=\sum_{x=-\infty}^\infty(c_x^\dagger c_{x+1}+c^\dagger_{x+1}c_x)\, ,
\end{equation}
where $c_x^\dagger,c_x$ are creation and annihilation operators at site 
$x$. The fermionic operators  obey standard canonical anticommutation 
relations. To diagonalize~\eqref{eq:ham} one defines a Fourier transform 
with respect to $x$, introducing the fermionic operators $b_k$ in momentum space 
as 
\begin{equation}
b_k:=\sum_{x=-\infty}^\infty e^{-ikx}c_x,\quad 
c_x=\int_{-\pi}^\pi \frac{dk}{2\pi} e^{i k x}b_k\,. 
\end{equation}
Eq.~\eqref{eq:ham} is rewritten in terms of $b_k$ as   
\begin{equation}
\label{eq:ham-k}
H=\int_{-\pi}^\pi\frac{dk}{2\pi} \varepsilon_k b^\dagger_k b_k\, ,\quad 
\varepsilon_k:=2\cos(k)\, .
\end{equation}
Eq.~\eqref{eq:ham-k} is diagonal, and it conserves the particle number. 
Let us consider a generic fermion density $n_f=k_F/\pi$, with $k_F$ the Fermi momentum.  
The ground state of~\eqref{eq:ham} is obtained by filling the single-particle states with  
quasimomenta $k$ in $k\in[-k_F,k_F]$.  The state with $n_f=1$ ($k_F=\pi$) in which all 
the quasimomenta are occupied is a product state, and it has trivial correlations. 
For intermediate filling $0<k_F<\pi$ the ground state of~\eqref{eq:ham} is critical, 
with power-law correlations. 

From the single-particle dispersion in~\eqref{eq:ham-k} we define 
the group velocity $v_k$ of the fermions as 
\begin{equation}
\label{eq:v-k}
v_k:=\frac{d\varepsilon_k}{dk}=-2\sin(k)\, . 
\end{equation}
Here we consider the out-of-equilibrium  dynamics under the Hamiltonian~\eqref{eq:ham} and 
localized loss processes at the center of the chain. These are treated in the formalism 
of quantum master equations~\cite{petruccione}. The time-evolved density matrix $\rho_t$ of the system 
is described by 
\begin{equation}
\label{eq:lind}
\frac{d\rho_t}{dt}=-i[H,\rho_t]+L^{-}\rho_t L^{-\, \dagger}-\frac{1}{2}\{L^{-\,\dagger} L^{-},\rho_t\}\, .
\end{equation}
Here, the so-called Lindblad jump operator $L^{-}$ is defined as  
$L^-=\sqrt{\gamma^-}c_0$ (see Fig.~\ref{fig0:cartoon} for a pictorial 
definition), with $\gamma^-$ the loss rate. 
Eq.~\eqref{eq:lind} describes incoherent absorption of fermions 
at the center of the chain. 

Entanglement properties of the systems can be extracted from the 
fermionic two-point correlation functions, i.e., the {\it covariance matrix} 
\begin{equation}
	G_{x,y}(t):=\mathrm{Tr}(c^\dagger_x c_y\rho(t))\, . 
\end{equation}
The  dynamics of $G_{x,y}$ is obtained as (we drop the dependence on the  
coordinates $x,y$ to lighten  the notation) 
\begin{equation}
\label{eq:G}
G(t)=
e^{t\Lambda}G(0)e^{t\Lambda^\dagger}, 
%+\int_0^t dze^{(t-z)\Lambda}\Gamma^+ e^{(t-z)\Lambda^\dagger}, 
\end{equation}
where $G(0)$ is the matrix containing the initial correlations. 
The matrix $\Lambda$ is defined as 
\begin{equation}
	\Lambda=ih-\frac{\Gamma^-}{2}, 
\end{equation}
where $h=\delta_{|x-y|,1}$ is the Hamiltonian contribution while 
$\Gamma^-=\gamma^-\delta_{x,0}$ encodes the localized dissipative effects.  
The covariance matrix $G_{x,y}$ is the solution of the 
linear system of equations  
\begin{equation}
\label{eq:one}
\frac{d G_{x,y}}{dt}=i(G_{x+1,y}+G_{x-1,y}-G_{x,y+1}-G_{x,y-1})
-\frac{\gamma^-}{2}(\delta_{x,0}G_{x,y}+\delta_{y,0}G_{x,y}). 
\end{equation}
Here we are interested in the hydrodynamic limit of large distances 
from the origin and long times, i.e., $x,y,t\to\infty$ with the 
ratios $x/t,y/t$ fixed. In this limit it can be shown  that 
the dissipation is effectively described by a delta potential. 
The strength of the potential is imaginary, which is a consequence of nonunitarity. 
Several properties of the system can be derived by studying the scattering problem 
of a quantum particle  with an imaginary delta potential~\cite{burke-2020}. For several 
initial states, both homogeneous as well as inhomogeneous ones,  
the dynamics of $G_{x,y}$ can be described solely in terms of the initial fermionic 
occupations and the reflection and transmission coefficients of the 
emergent delta potential~\cite{alba2021noninteracting}. Here we are interested 
in the situation in which the initial state of the dynamics 
is a Fermi sea with arbitrary Fermi momentum $k_F$. 

%%%%%%%%%%%%%%%%%%%%%%%%%%%%%%%%%%%%%%%%%%%%%%%%%%%%%%%%%%%%%%%%%%%%%%%%%%%%%%%%%%
\subsection{Hydrodynamic limit of the covariance matrix}
\label{sec:cov-hydro}

In the hydrodynamic limit the solution of~\eqref{eq:one} with initial condition the 
Fermi sea is obtained as~\cite{alba2021noninteracting} (see also~\cite{froml-2020})
\begin{equation}
\label{eq:f-corr}
G_{x,y}(t)=\int_{-k_F}^{k_F}\frac{dk}{2\pi}(e^{ikx}+\chi_{x}(t)r(k)e^{i|kx|})
(e^{-iky}+\chi_{y}(t)r(k)e^{-i|ky|}). 
\end{equation}
Notice the absolute values in the second terms in the brackets. Moreover, one 
should observe that the contributions associated with the two coordinates $x,y$ 
factorize. This factorization is crucial~\cite{krapivsky-2019} to 
obtain the exact solution of~\eqref{eq:one}. 
In~\eqref{eq:f-corr} $r(k)$ is the momentum-dependent reflection amplitude of the effective delta 
potential describing the dissipation source at the origin. 
The analytic expression for $r(k)$ and for the associated transmission amplitude 
$\tau(k)$ are given as~\cite{alba2021noninteracting} 
\begin{equation}
	\label{eq:delta-coeff}
	r(k):=-\frac{\gamma^-}{2}\frac{1}{\frac{\gamma^-}{2}+|v_k|},\quad
	\tau(k):=\frac{|v_k|}{\frac{\gamma^-}{2}+|v_k|}, 
\end{equation}
where $v_k$ is the fermion group velocity defined in~\eqref{eq:v-k}. 
Notice that~\eqref{eq:delta-coeff} coincide with the reflection and transmission 
amplitude for a quantum particle scattering with a delta potential with imaginary 
strength $-i\gamma^-/2$ after redefining~\cite{burke-2020} $v_k\sim k$. 
Crucially, since the dynamics is nonunitary one has that 
\begin{equation}
\label{eq:abs}
|a|^2:=1-|r|^2-|\tau|^2=\frac{\gamma^- |v_k|}{(\frac{\gamma^-}{2}+|v_k|)^2}>0, 
\end{equation}
where we defined the absorption coefficient $|a|^2$, which is the 
probability that a fermion with quasimomentum $k$ is removed at the 
origin. 

The time dependence of the correlator in~\eqref{eq:f-corr} is encoded in the function 
$\chi_x$, which is defined as 
\begin{equation}
	\label{eq:constr}
	\chi_{x}:=\Theta(|v_k|t-|x|). 
\end{equation}
At $t=0$ from~\eqref{eq:f-corr} one recovers the initial correlation of 
the Fermi sea as 
\begin{equation}
	\label{eq:fs-corr}
	G_{x,y}(0)=\frac{\sin(k_F(x-y))}{\pi(x-y)}. 
\end{equation}
%

%
%########################################
\begin{figure}[t]
\begin{center}
\includegraphics[width=0.5\textwidth]{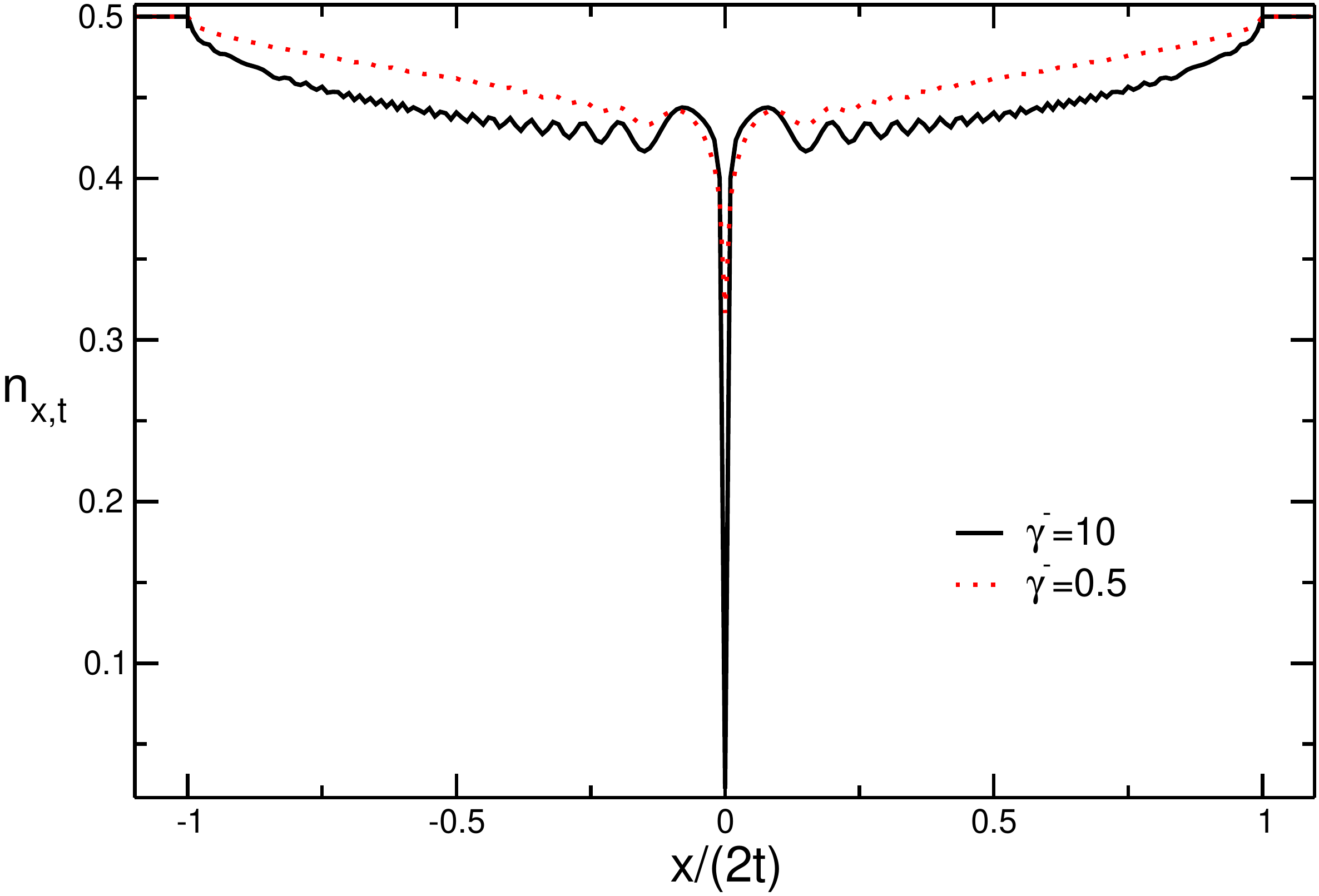}
\caption{Dynamics of the fermionic density $n_{x,t}$ in the presence of 
	localized losses. Results are for the initial Fermi sea 
	with $k_F=\pi/2$ and for loss rate $\gamma^-=10$ and 
	$\gamma^-=0.5$ (continuous and dotted lines, respectively). 
	The oscillations are an artifact of the 
	approximations and vanish in the hydrodynamic limit 
	$x,t\to\infty$ with their ratio fixed. Notice that in the 
	hydrodynamic limit the density develops a 
	discontinuity at the origin. 
}
\label{fig1aa:density}
\end{center}
\end{figure}
%########################################
%
To get an idea of the effect of the dissipation, it is instructive to consider 
the dynamics of the local fermionic density $n_{x,t}$ 
\begin{equation}
	n_{x,t}=G_{x,x}. 
\end{equation}
This is discussed in Fig.~\ref{fig1aa:density}. We plot $n_{x,t}$ versus 
the scaling variable $x/(2t)$, showing results for $\gamma^-=0.5$ and $\gamma^-=10$. 
The results are obtained by using~\eqref{eq:f-corr}. 
We focus on the effects of the localized losses on the initial Fermi sea 
with $k_F=\pi/2$. As expected, the Fermi seas gets depleted with time and 
a nontrivial density profile forms around the origin. For $|x/(2t)|>1$ the 
effect of the dissipation is not present and one has the initial density $1/2$. 
The density profile exhibits a discontinuity at the origin. This 
reflects the presence of an 
effective delta potential at the origin. Finally,  
the oscillations present in Fig.~\ref{fig1aa:density} are an artifact of the 
approximations employed to derive~\eqref{eq:f-corr}, and vanish in the 
hydrodynamic limit. In the strong 
dissipation limit $\gamma^-\to\infty$ the evolution of the density freezes. 
This is a manifestation of the quantum Zeno effect~\cite{degasperis-1974,misra-1977,facchi-2002}. 
In the following sections we show that the fermionic dynamics shown 
in Fig.~\ref{fig1aa:density} is accompanied with a robust linear entanglement 
growth with time.

%#############################################################
\section{Entanglement entropies and logarithmic negativity: Definitions}
\label{sec:obs}

In order to understand how the presence of localized losses affects the entanglement 
content of the system here we focus on several quantum-information-related quantities, 
such as the entanglement entropies and the logarithmic negativity. 
To introduce them, let us consider a bipartition of the system as $A\cup \bar A$ (see, for 
instance, Fig.~\ref{fig0:cartoon} (a) and (b)). By tracing over the  degrees 
of freedom of $\bar A$, which is the complement of $A$, 
one obtains the reduced density matrix 
$\rho_A=\mathrm{Tr}_{\bar A}\rho$,  where 
$\rho$ is the full-system density matrix. 
The R\'enyi entropies are defined as~\cite{amico2008entanglement,eisert2010colloquium,calabrese2009entanglemententropy,laflorencie2016quantum}  
\begin{equation}
	\label{eq:renyi-def}
	S^{(n)}:=\frac{1}{1-n}\mathrm{Tr}(\rho^n_A),\quad\mathrm{with}\, n\in\mathbb{R}. 
\end{equation}
In the limit $n\to1$ one obtains the von Neumann entropy as 
\begin{equation}
\label{eq:vn}
	S=-\mathrm{Tr}\rho_A\ln(\rho_A). 
\end{equation}
Both R\'enyi and von Neumann entropies are good entanglement measures 
provided that the full system is in a pure state. However, in the 
presence of dissipation the full system is 
in a mixed state, which  introduces some ``classical'' correlation between 
$A$ and $\bar A$. This spurious, i.e., non-quantum, correlation,  
affects both the R\'enyi entropies and the entanglement entropy. 

In these situations the so-called logarithmic negativity~\cite{lee2000partial,vidal2002computable} 
can be used to quantify the amount of genuine entanglement between $A$ and the rest. 
The logarithmic negativity ${\cal E}$ is defined from the partially-transposed 
density matrix $\rho^{T}$. This is defined from $\rho$ by taking the matrix 
transposition with respect to the degrees of freedom of $\bar A$ as 
\begin{equation}
	\langle e_i,\bar e_j|\rho^T| e_k,\bar e_l\rangle=
	\langle e_i,\bar e_l|\rho|e_i,\bar e_j\rangle, 
\end{equation}
with $e_i,\bar e_j$ two bases for $A$ and its complement, respectively. 
Unlike $\rho$, $\rho^T$ is not positive definite, and its negative 
eigenvalues quantify the amount of entanglement. The logarithmic negativity is 
defined as 
\begin{equation}
	\label{eq:neg-def0}
	{\cal E}=\ln(\mathrm{Tr}|\rho^T|). 
\end{equation}
Here we focus on free-fermion models. 
For free-fermion and free-boson models both the R\'enyi entropies and 
the von Neumann entropy of a region $A$ are calculable from the fermionic correlation 
function $G_{x,y}$ restricted to $A$, i.e., with $x,y\in A$. 
Specifically, the R\'enyi entropies are obtained as~\cite{peschel2009reduced}  
\begin{equation}
	\label{eq:renyi-def}
	S^{(n)}=\frac{1}{1-n}\mathrm{Tr}\ln\Big[G^n+(1-G)^n\Big]. 
\end{equation}
In the limit $n\to1$, one recovers the von Neumann entropy as 
\begin{equation}
	\label{eq:vn-def}
	S=-\mathrm{Tr}(G\ln(G)-(1-G)\ln(1-G)). 
\end{equation}
The logarithmic negativity ${\cal E}$ can be calculated efficiently 
from the two-point function only for free bosons~\cite{audenaert2002entanglement}. 
For free fermions the partial transposed $\rho^T$ is not a gaussian operator, 
although it can be written as a sum of two gaussian operators~\cite{eisler2015partial} 
as 
\begin{equation}
	\label{eq:opm}
	\rho^T=e^{-i\pi/4} O_++e^{i\pi/4}O_-, 
\end{equation}
where $O_\pm$ are gaussian operators. Very recently, 
an alternative definition of negativity 
has been put forward~\cite{shapourian2017many,shapourian2017partial,shapourian2019entanglement}. 
We dub this alternative negativity  {\it fermionic} negativity. Its definition 
reads as 
\begin{equation}
	\label{eq:e-f}
	{\cal E}:=\ln\mathrm{Tr}\sqrt{O_+O_-}. 
\end{equation}
Here we use the same symbol ${\cal E}$ for the fermionic negativity and for 
the standard one in~\eqref{eq:neg-def0} because in the following we will only 
use the fermionic one. 
In contrast with~\eqref{eq:neg-def0}, since the product $O_+O_-$ is a gaussian 
operator, the fermionic negativity~\eqref{eq:e-f} can be computed 
effectively in terms of fermionic two-point functions. 
Specifically, let us rewrite the full-system correlation matrix $G$ 
as  
\begin{equation}
	G=\left(\begin{array}{cc} 
		G_{AA} & G_{A\bar A}\\
		G_{\bar A A} & G_{\bar A\bar A}
\end{array}\right)
\end{equation}
Here $G_{WZ}$, with $W,Z=A,\bar A$ is obtained from the 
full system $G_{x,y}$ restricting to $x\in W$ and $y\in Z$. 
One now defines the matrices $G^\pm$ as
\begin{equation}
	G^\pm=\left(\begin{array}{cc}
			-G_{AA} & \pm i G_{A\bar A}\\
			\pm i G_{\bar A A} & G_{\bar A\bar A}
		\end{array}
\right)
\end{equation}
We then define the matrix $C^T$ as  
\begin{equation}
	C^T=\frac{1}{2}\mathbb{I}-P^{-1}(G^++G^-),\quad\mathrm{with}\, 	P=\mathbb{I}-G^+G^-. 
\end{equation}
From the eigenvalues $\xi_i$ of $C^T$ and $\lambda_i$ of $G$ we can define the 
fermionic negativity ${\cal E}$ as~\cite{shapourian2017partial} 
\begin{equation}
	\label{eq:neg-def}
	{\cal E}=\sum_i\Big[\ln[\xi_i^{1/2}+(1-\xi_i)^{1/2}]+\frac{1}{2}\ln[\lambda_i^2+(1-\lambda_i)^2]\Big], 
\end{equation}
It has been shown in Ref.~\cite{shapourian2019entanglement} that under 
reasonable assumptions the fermionic negativity is a good entanglement measure for 
mixed states. 

%#############################################################
\section{Hydrodynamic description of entanglement entropies}
\label{sec:hydro-ent}

We now discuss the out-of-equilibrium dynamics of the entanglement 
entropies in the hydrodynamic limit. Before that,  
we provide a more general result, which allows us to obtain the hydrodynamic 
behavior of the trace of a generic function of the fermionic correlator 
(cf.~\eqref{eq:f-corr}).  
Let us  consider the bipartitions in Fig.~\ref{fig0:cartoon} (a) and 
(b). In Fig.~\ref{fig0:cartoon} (a) subsystem $A$ is the interval $[0,\ell]$, i.e., 
on the right of the dissipative impurity. In Fig.~\ref{fig0:cartoon} (b) we 
consider subsystem $A'=[-\ell/2,\ell/2]$ centered around the impurity. 
Let us consider a generic function ${\mathcal F}(z)$, and let us focus on the 
quantity $\mathrm{Tr}{\mathcal F}(G_X)$, with $X=A,A'$. 
In the hydrodynamic limit $t,\ell\to\infty$, with their ratio fixed, 
one can show that 
\begin{multline}
	\label{eq:F}
	\mathrm{Tr}{\mathcal F}(G_X)=\ell\int_{-k_F}^{k_F}\frac{dk}{2\pi}
	\Big[\Big(1-\frac{1}{2 z_X}\min(z_X|v_k|t/\ell,1)\Big){\mathcal F}(1)
\\+\frac{1}{2 z_X}{\mathcal F}(1-z_X|a(k)|^2)
	\min(z_X|v_k|t/\ell,1)\Big],\quad z_{A(A')}=1(2). 
\end{multline}
Here $v_k$ is the fermion group velocity in~\eqref{eq:v-k}, and $|a(k)|^2$ 
is the absorption coefficient of the emergent delta potential (cf.~\eqref{eq:abs})  
at the origin. 
Eq.~\eqref{eq:F} depends only on $1-|a|^2$, i.e., the probability  of the 
fermions not to be absorbed at the origin. Also, the only 
dependence on time is via the factor $\min(z_X|v_k|t/\ell)$, which 
encodes the fact that $A$ and $A'$ are finite, and information propagates from 
the origin at a finite velocity $v_k$. The factor $z_X$ in~\eqref{eq:F} accounts for 
the different geometries in Fig.~\ref{fig0:cartoon} (a) and (b), and it has a simple 
interpretation. For instance, in the argument of the second term in~\eqref{eq:F} 
$z_X$ takes into account that for the bipartition in Fig.~\ref{fig0:cartoon} (a) 
the number of absorbed fermions is twice that for the bipartition in Fig.~\ref{fig0:cartoon} (a) 
because the impurity is at the center of $A'$. Moreover, in 
$\min(z_X |v_k|t/\ell,1)$, $z_X$  reflects that for $A'$ the 
distance between the impurity and the edge of $A'$ is $\ell/2$ instead 
of $\ell$. 

For generic ${\mathcal F}(z)$, Eq.~\eqref{eq:F} predicts a linear behavior 
with time for $t\le \ell/(z_X v_\mathrm{max})$, with $v_\mathrm{max}$ the 
maximum velocity in the system. This is followed by an asymptotic saturation at $t\to\infty$ to a 
volume law $\propto\ell$. 
Finally, for $\gamma^-=0$ one recovers the unitary case and 
from~\eqref{eq:F}, one obtains that 
\begin{equation}
	\label{eq:nodiss}
	\mathrm{Tr}{\mathcal F}(G_X)=\ell\int_{-kF}^{k_F}\frac{dk}{2\pi} {\mathcal F}(1). 
\end{equation}
Eq.~\eqref{eq:nodiss} means that in the absence of dissipation there is no dynamics 
and for any ${\mathcal F}$ one has a constant contribution that is proportional to $\ell$. 
The fact that there is no dependence on $z_X$ and on the geometry 
reflects translation invariance. 

The derivation of~\eqref{eq:F} is reported in Appendix~\ref{sec:app} and 
it relies on the multidimensional stationary phase approximation~\cite{wong}, and  
on the assumption that ${\mathcal F}(z)$ admits a Taylor expansion around $z=0$. We should 
also stress that although we discuss only the two geometries in Fig.~\ref{fig0:cartoon} 
(a) and (b), it should be possible to generalize~\eqref{eq:F} to arbitrary bipartitions or 
multipartitions. In the following, by considering different functions ${\mathcal F}(z)$ 
we provide exact results for the moments of the correlation matrix and the entanglement 
entropies in the hydrodynamic limit. 

%
%########################################
\begin{figure}[t]
\begin{center}
\includegraphics[width=0.5\textwidth]{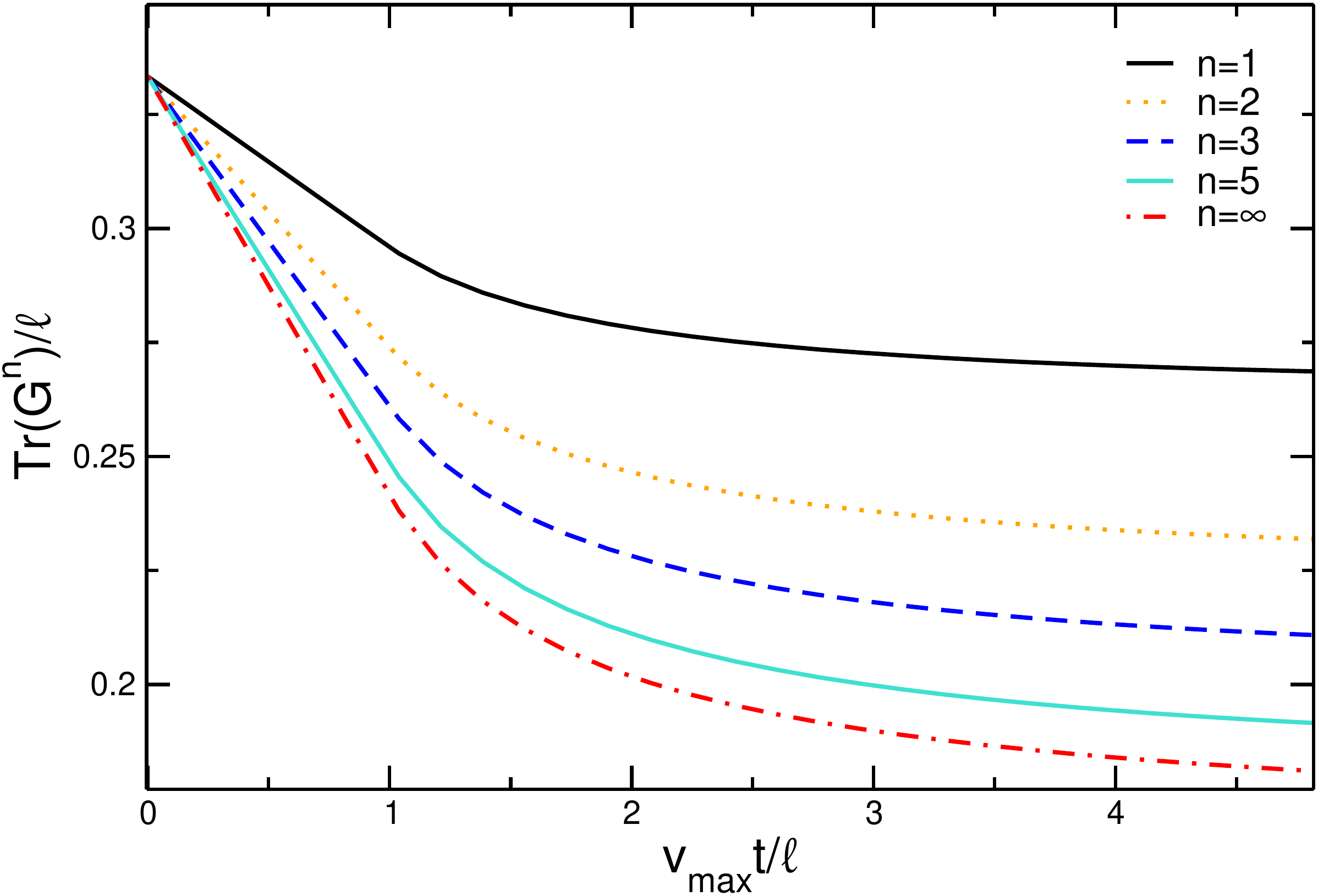}
\caption{ Moments of the fermionic correlator $M_n=\mathrm{Tr}(G^n)$, where 
 $G$ is the fermionic correlation function restricted to subsystem $A$ 
 (see Fig~\ref{fig0:cartoon}). We plot the rescaled moments $M_n/\ell$, 
 with $\ell$ the length of $A$ versus $v_\mathrm{max} t/\ell$, $v_\mathrm{max}$  being the 
 maximum velocity. Lines are analytic results in the hydrodynamic limit 
 $\ell,t\to\infty$ with their ratio fixed. We only show results for $\gamma^-=1$ and $k_F=\pi/3$.  
 Note the linear behavior for $t\le \ell/v_M$ followed by a saturation 
 at $t\to\infty$. 
}
\label{fig1:traces-theo}
\end{center}
\end{figure}
%########################################
%

%#############################################################
\subsection{Moments of the correlation matrix}
\label{sec:mn}

Here we study the hydrodynamic limit of the moments $M_n$ of the fermionic 
correlation matrix. These are defined as 
\begin{equation}
	M_n=\mathrm{Tr}(G^n),  
\end{equation}
where the  correlation matrix $G$ is restricted to subsystem 
$A,A'$ (see Fig.~\ref{fig0:cartoon}). 
The behavior of $M_n$ in the hydrodynamic limit is readily obtained 
from~\eqref{eq:F} by choosing ${\mathcal F}(z)=z^n$. 
One obtains that 
\begin{multline}
	\label{eq:Mn}
	M_n=\ell\int_{-k_F}^{k_F}\frac{dk}{2\pi}
	\Big[\Big(1-\frac{1}{2 z_X}\min(z_X|v_k|t/\ell,1)\Big)
\\+\frac{1}{2 z_X}(1-z_X|a(k)|^2)^n
	\min(z_X|v_k|t/\ell,1)\Big],\quad z_{A(A')}=1(2). 
\end{multline}
The structure is the same as in~\eqref{eq:F}. 
$M_n$ exhibit the same qualitative behavior with a linear decrease 
at short times $t\le\ell/v_\mathrm{max}$, which is 
followed by an asymptotic saturation at $t\to\infty$. 
Several remarks are in order. 
First, at $t=0$ one has that for any $n$, $M_n=\ell k_F/\pi$, which is 
the initial number of fermions in the subsystem. 
For $t\to\infty$ one has that the number of fermions $M_1$ in the subsystem 
is 
\begin{equation}
	M_1 \xrightarrow{t\to\infty}\ell\int_{-k_F}^{k_F}\frac{dk}{2\pi}\Big(1-\frac{|a|^2}{2}\Big).
\end{equation}
This means that  $M_1\propto\ell$ for $t\to\infty$, despite the presence of dissipation. 
In the strong dissipation limit  $\gamma^-\to\infty$ one 
has that $|a|^2\to0$, and $M_1\to\ell k_F/\pi$, i.e., the initial fermion number. 
This is a manifestation of the quantum Zeno effect. In the limit $\gamma^-\to\infty$ 
the dynamics of the system is arrested and the number of fermions absorbed 
at the origin vanishes. 
Finally, it is interesting to consider $M_n$ in 
the limit  $n\to\infty$. One can readily check that $1-z|a|^2<1$, 
which implies that only the first term in~\eqref{eq:Mn} 
survives. In particular, in the limit $t\to\infty$, from~\eqref{eq:Mn} one 
obtains that $M_\infty=\ell k_F/\pi(1-1/(2z_X))$. For $z_X=1$ (i.e., for 
the partition in Fig.~\ref{fig0:cartoon} (a)) one has $M_\infty=\ell k_F/(2\pi)$, 
which is half of the initial number of fermions. 

In Fig.~\ref{fig1:traces-theo} we show numerical predictions for 
$M_n$ obtained by using~\eqref{eq:Mn}. We consider the case with $k_F=\pi/3$ and 
we restrict ourselves to $\gamma^-=1$. We provide results only for the 
bipartition in Fig.~\ref{fig0:cartoon} (a). The generic behavior outlined above 
is clearly visible in the figure. 

%
%########################################
\begin{figure}[t]
	\begin{center}
	\includegraphics[width=0.5\textwidth]{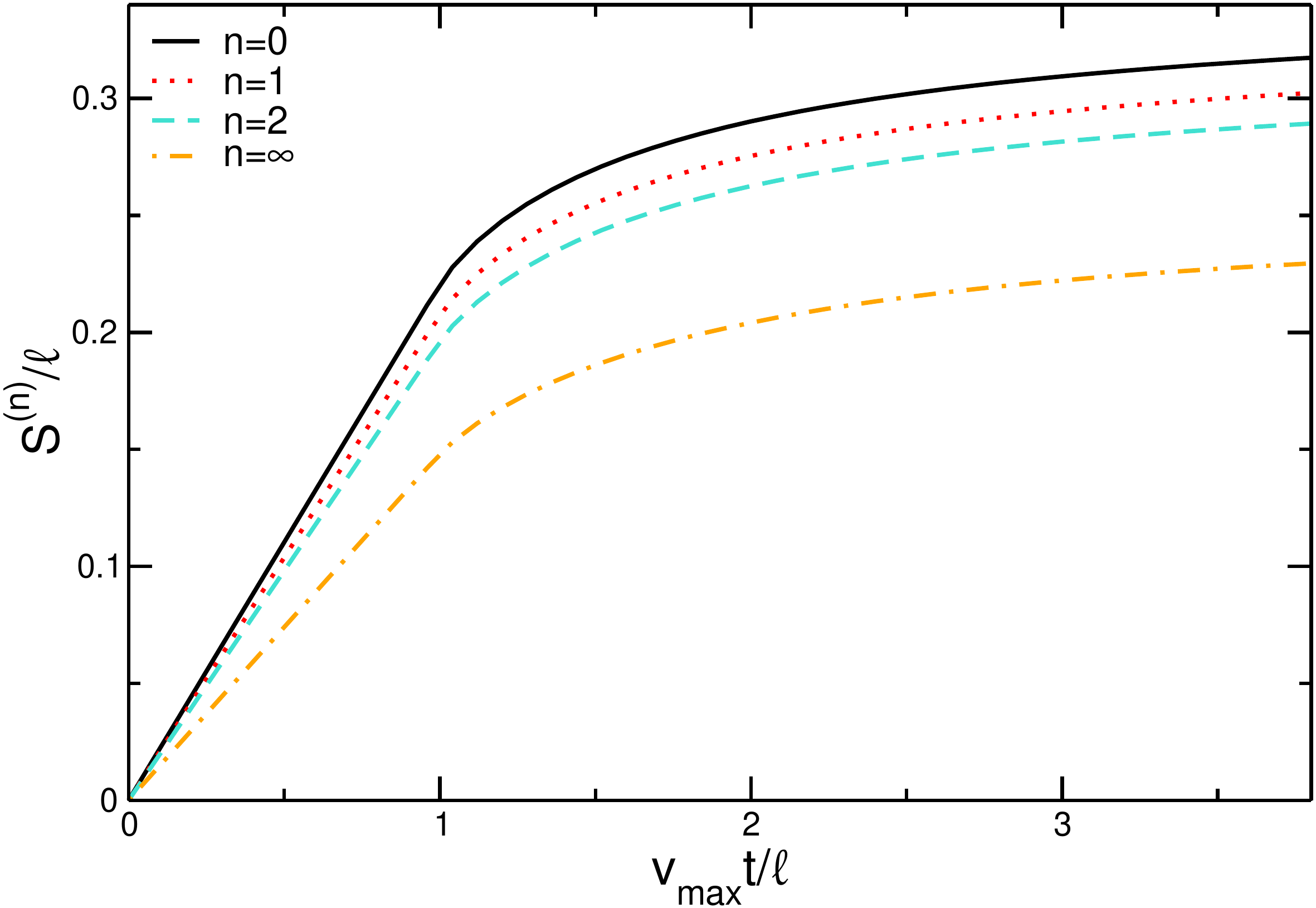}
\caption{ Entanglement entropies $S^{(n)}$ of a subsystem $A$ 
 placed next to the dissipation source (see Fig.~\ref{fig0:cartoon} (a)).
 The different lines are analytic predictions in the hydrodynamic 
 limit for different values of $n$. We plot $S^{(n)}/\ell$ versus 
 $v_\mathrm{max}t/\ell$, with $\ell$ the size of $A$ and $v_\mathrm{max}$ 
 the maximum velocity. We only show results for $\gamma^-=1$ and $k_F=\pi$. 
}
\label{fig2:ent-theo}
\end{center}
\end{figure}
%########################################
%

%#############################################################
\subsection{Entanglement entropies}
\label{sec:ent-theo}

The hydrodynamic limit of the entanglement entropies, both the von Neumann and the 
R\'enyi entropies, is  obtained from~\eqref{eq:F} by choosing 
\begin{equation}
	\label{eq:Hn}
	{\mathcal F}(z)=H_n(z)=\frac{1}{1-n}\ln[z^n+(1-z)^n]. 
\end{equation}
In the limit $n\to1$ one recovers the von Neumann entropy 
by choosing $H_1(z)=-z\ln(z)-(1-z)\ln(1-z)$. After using~\eqref{eq:Hn} in~\eqref{eq:F}, 
and after observing that for any $n$, ${\mathcal F}(1)=0$, one obtains that 
\begin{equation}
	\label{eq:ent-hydro}
	S^{(n)}=\frac{1}{2 z_X}\frac{\ell}{1-n}\int_{-k_F}^{k_F}\frac{dk}{2\pi} H_n(1-z_X|a|^2)\min(z_X|v_k|t/\ell,1). 
\end{equation}
First, for $\gamma^-=0$, i.e., in the absence of dissipation, one has 
that $S^{(n)}=0$ for any $n$. This is consistent with the fact that 
for a Fermi sea the entanglement entropies exhibit 
the typical Conformal Field Theory (CFT)  logarithmic scaling 
as~\cite{calabrese2009entanglemententropy}  
\begin{equation}
	\label{eq:cft-scaling}
	S^{(n)}= \frac{c}{6}\Big(1+\frac{1}{n}\Big)\ln(\ell)+c_n, 
\end{equation}
where $c=1$ is the central charge of the model and $c_n$ are nonuniversal 
constants. The scaling~\eqref{eq:cft-scaling} cannot be captured 
by~\eqref{eq:ent-hydro}, which describes the leading volume-law behavior $S^{(n)}\propto\ell$. 
In the strong dissipation limit $\gamma^-\to\infty$, one 
has that, reflecting the Zeno effect, $S^{(n)}$ vanish for any $n$. 

Away from the limits $\gamma^-=0$ and $\gamma^-\to\infty$,  the entanglement 
entropies increase linearly at short times $t\le \ell/(z_X v_\mathrm{max})$, and  
saturate to a volume-law scaling $S^{(n)}\propto\ell$ 
at asymptotically long times. It is interesting to consider the limit 
$n\to\infty$, which gives the so-called single-copy entanglement. 
From~\eqref{eq:ent-hydro} it is clear that only the  first term inside the 
logarithm in~\eqref{eq:Hn} counts, and one obtains that 
\begin{equation}
	\label{eq:s-copy}
	S^{(\infty)}=-\frac{\ell}{2 z_X}\int_{-k_F}^{k_F}\frac{dk}{2\pi} \ln(1-z_X|a|^2)\min(z_X|v_k|t/\ell,1). 
\end{equation}
It is now  crucial to remark that Eq.~\eqref{eq:ent-hydro} gives the same qualitative 
behavior for the entanglement entropies of $A$ and $A'$ (see Fig.~\ref{fig0:cartoon} (a) and (b)). 
This is surprising at first because no production of entanglement is expected for the centered 
bipartition in Fig.~\ref{fig0:cartoon} (b). The reason is that the reflected and the 
transmitted fermions, which form the entangled pairs, are never shared between $A'$ 
and its complement. 
The linear growth in this case  should be attributed to the 
formation of a nontrivial density profile around the origin, which  reflects the 
creation of thermodynamic entropy. The entanglement entropies are not {\it bona fide} 
entanglement measures for mixed states because they are sensitive to this 
thermodynamic contribution. We anticipate that, in contrast, the 
logarithmic negativity is sensitive to the genuine quantum correlation only 
(see section~\ref{sec:num-b}). 

In Fig.~\ref{fig2:ent-theo} we report analytic predictions for the 
dynamics of the entanglement entropies obtained from~\eqref{eq:ent-hydro}. 
We plot the rescaled entropies $S^{(n)}_X/\ell$ versus $v_\mathrm{max}t/\ell$ for 
several values of $n$. We consider only the bipartition in Fig.~\ref{fig0:cartoon} 
(a), i.e., we choose $X=A$ in~\eqref{eq:ent-hydro}. Furthermore, we show data 
for $k_F=\pi$ and $\gamma^-=1$. The qualitative behaviour discussed above is clearly visible. 

%
%########################################
\begin{figure}[t]
\begin{center}
\includegraphics[width=0.5\textwidth]{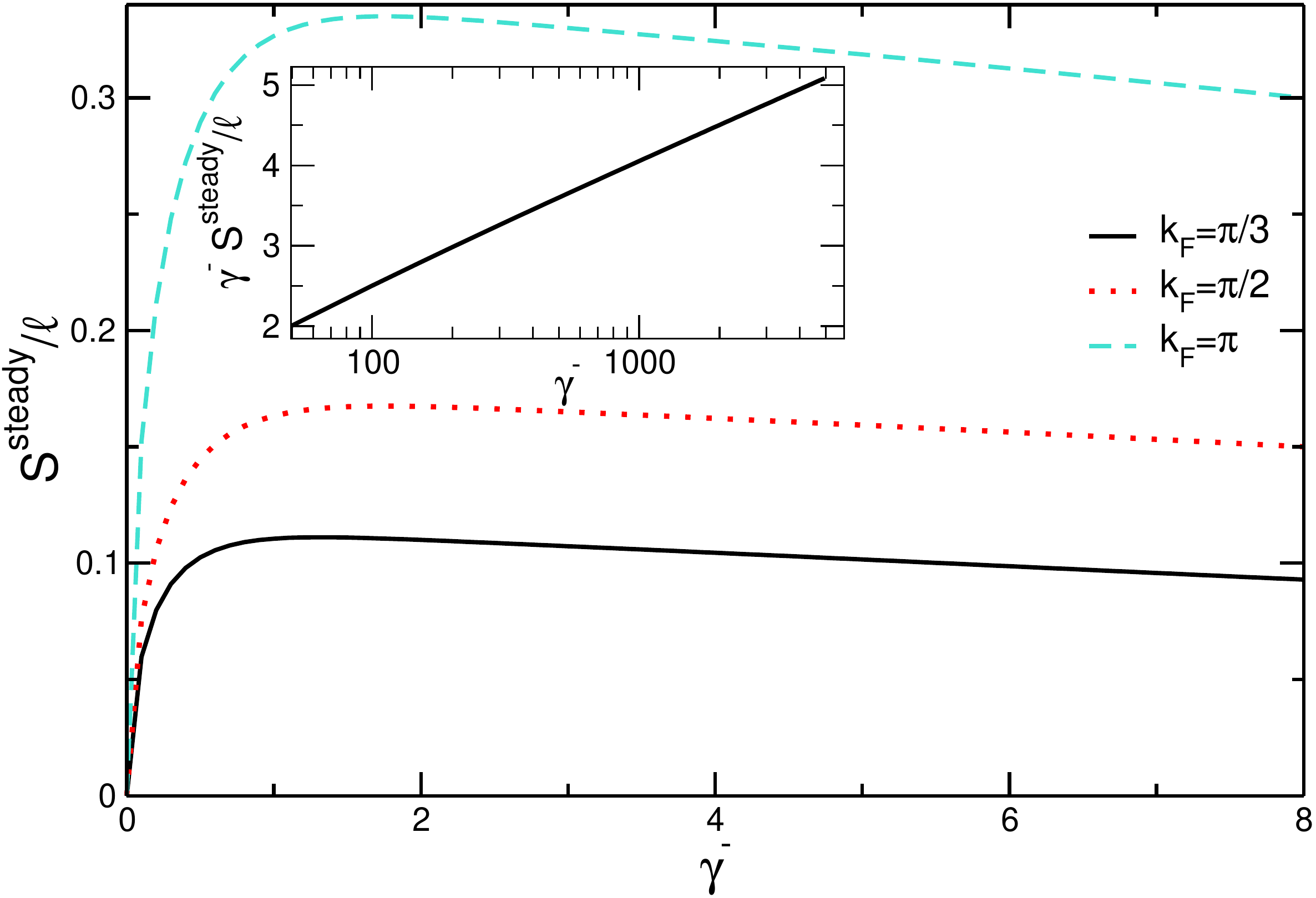}
\caption{ Steady-state entropy in the free fermion chain with 
 localized losses. We show results for the bipartition in Fig.~\ref{fig0:cartoon} (a). 
 We plot $S^{(\mathrm{steady})}/\ell$ versus the loss 
 rate $\gamma^-$. The different lines in the main figure are for 
 initial states with different Fermi momentum $k_F$. Note that 
 the steady-state entropy has a maximum at $\gamma^-\approx 1$. 
 For $\gamma^-\to\infty$ the steady-state entropy vanishes as 
 $S^{(\mathrm{steady})}/\ell\propto\ln(\gamma^-)/\gamma^-$, as it is 
 shown in the inset. 
}
\label{fig3:ent-max}
\end{center}
\end{figure}
%########################################
%

%
%########################################
\begin{figure}[t]
\begin{center}
\includegraphics[width=0.5\textwidth]{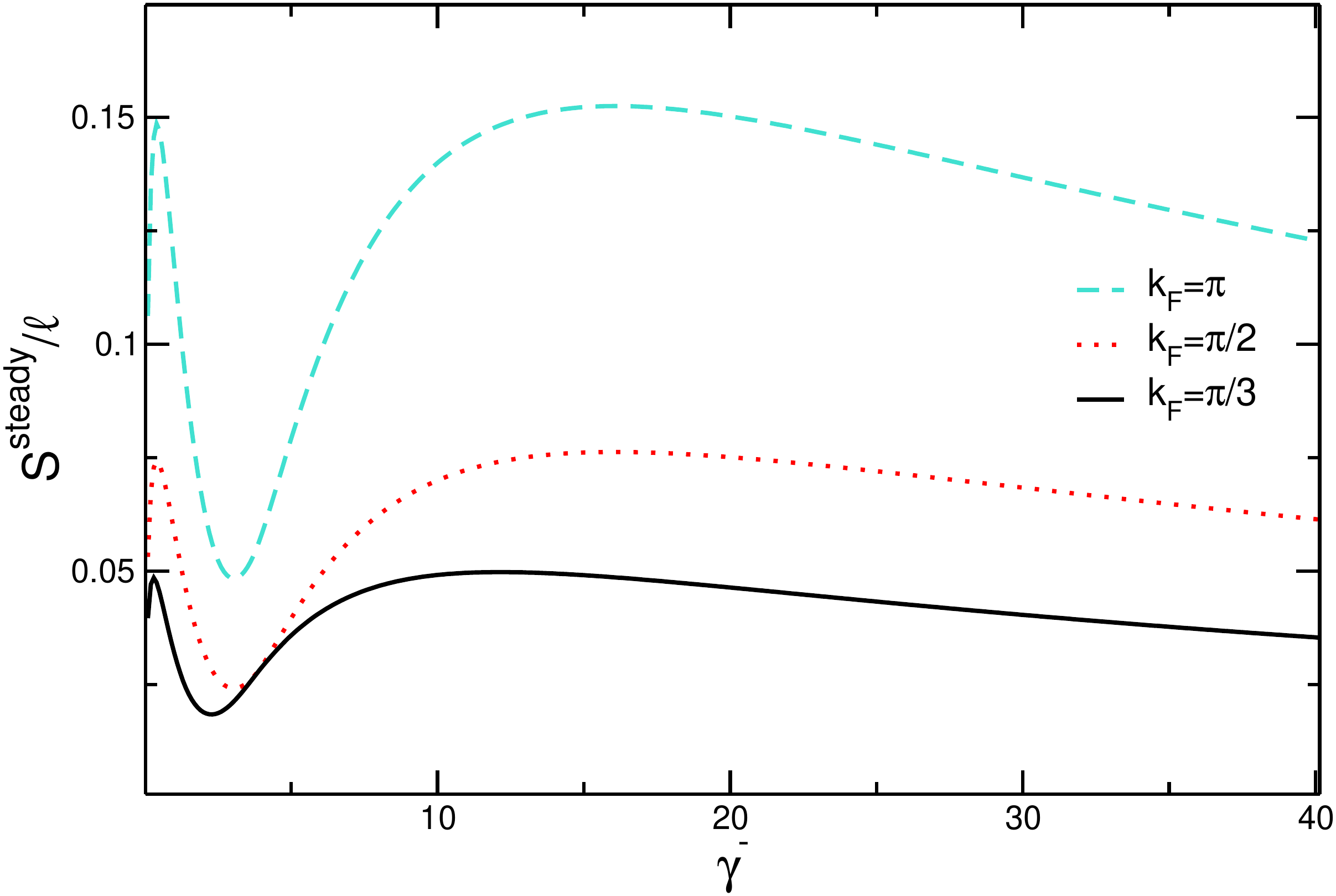}
\caption{ Same as in Fig.~\ref{fig3:ent-max} for the 
 centered partition in Fig.~\ref{fig0:cartoon} (b). 
 Notice the presence of two maxima, in contrast 
 with Fig.~\ref{fig3:ent-max}. 
}
\label{fig3a:ent-max}
\end{center}
\end{figure}
%########################################
%

%#############################################################
\subsection{Zeno death of entanglement entropy}
\label{sec:ss-entropy}

It is interesting to investigate the steady-state value of the 
entanglement entropy as a function of the dissipation rate $\gamma^-$. 
The steady-state entanglement entropy $S^{\mathrm{steady}}$ is 
obtained from~\eqref{eq:ent-hydro} as 
\begin{equation}
	\label{eq:s-steady}
	S^{(\mathrm{steady})}=\frac{\ell}{2z_X}\int_{-k_F}^{k_F}\frac{dk}{2\pi} 
	H_1(1-z_X|a|^2). 
\end{equation}
In Fig.~\ref{fig3:ent-max} we plot $S^{(\mathrm{steady})}/\ell$ 
versus $\gamma^-$. The results are for $X=A$ (see Fig.~\ref{fig0:cartoon} (a)). 
In the main plot, the different curves correspond to different values of $k_F$. 
Notice that the entanglement entropy increases upon increasing 
$k_F$. This is expected because the entanglement entropy is proportional 
to the number of fermions that scatter with the impurity at the origin. 
Interestingly, the data exhibit a maximum in the 
region $\gamma^-\in[1.5,2]$. In the strong dissipation limit $\gamma^-\to\infty$ 
the entanglement entropy vanishes. This is a consequence of the quantum 
Zeno effect. The decay is as $S^{(\mathrm{steady})}
\propto \ln(\gamma^-)/\gamma^-$ (see the inset of Fig.~\ref{fig3:ent-max}).  

Finally, it is interesting to compare with the result  
for the centered partition in Fig.~\ref{fig0:cartoon} (b). 
This is discussed in Fig.~\ref{fig3a:ent-max}. 
A richer structure is observed. Indeed, the steady-state entropy exhibits two maxima, one  
at ``weak'' dissipation for $\gamma^-\approx 0.5$ and one 
in the ``strong'' dissipation regime for $\gamma^-\approx 10$. 
Notice also that the steady-state entropy is generically 
smaller than in Fig.~\ref{fig3:ent-max}.

%#############################################################
\section{Numerical benchmarks}
\label{sec:num}

We now provide numerical benchmarks for the results derived in 
section~\ref{sec:hydro-ent}. We discuss the moments $M_n$ (cf~\eqref{eq:Mn}) 
in section~\ref{sec:num-a}. In section~\ref{sec:num-b} we focus on the entanglement 
entropies. Importantly, we discuss the interplay between entanglement 
and thermodynamic correlation  by comparing the evolution of the von 
Neumann entropy and that of the logarithmic negativity for the two 
bipartitions in Fig.~\ref{fig0:cartoon} (a) and (b).

%
%########################################
\begin{figure}[t]
\begin{center}
\includegraphics[width=0.9\textwidth]{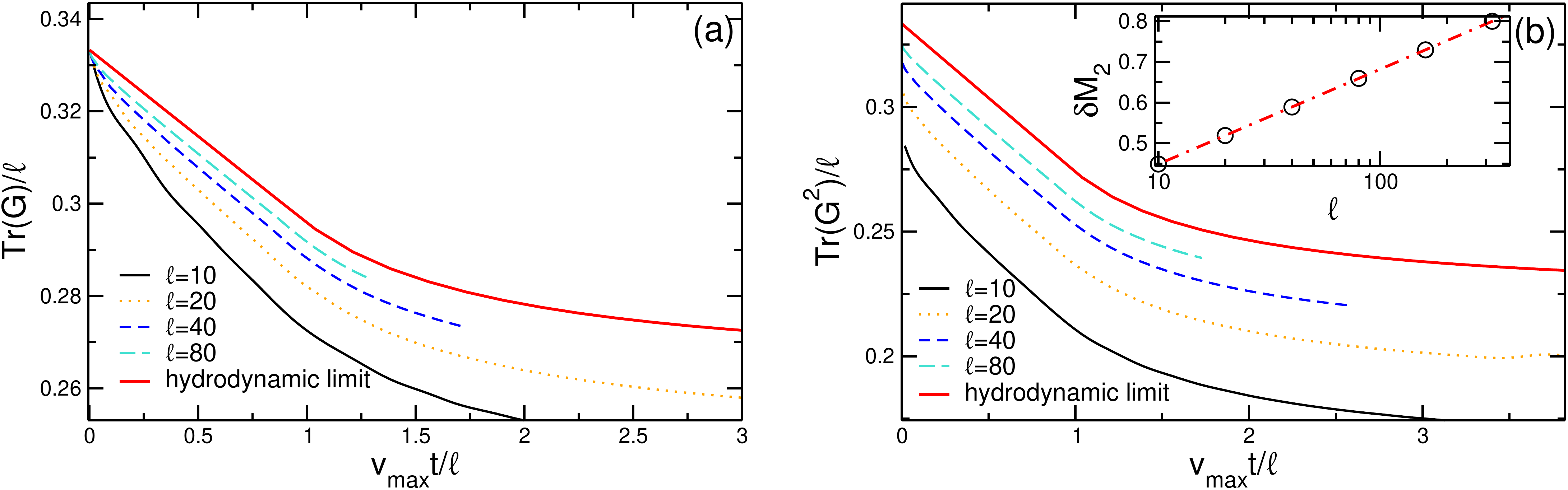}
\caption{ Moments of the fermionic correlator $M_n=\mathrm{Tr}(G^n)$ 
 restricted to subsystem $A$ (bipartition in Fig~\ref{fig0:cartoon} (a)). 
 We show the rescaled moments $M_n/\ell$, with $\ell$ the 
 size of $A$ plotted versus $v_\mathrm{max} t/\ell$. Here $v_\mathrm{max}$ 
 is the maximum velocity. All the results are for $\gamma^-=1$. The two panels are for 
 $n=1$ and $n=2$. Different lines denote different subsystem size $\ell$. 
 The dashed-dotted line is the analytic result in the hydrodynamic limit. 
 Sizeable finite-time and finite-size corrections are present. In (b) we 
 show the deviation from the hydrodynamic result at $t=0$, 
 $\delta M_2=M_2^{\mathrm{hydro}}-M_2$ as a function of 
 $\ell$. Notice the logarithmic scale on the $x$-axis. 
}
\label{fig4:traces-check}
\end{center}
\end{figure}
%########################################
%

%############################################################
\subsection{Moments of fermionic correlators}
\label{sec:num-a}

%
%########################################
\begin{figure}[t]
\begin{center}
\includegraphics[width=0.5\textwidth]{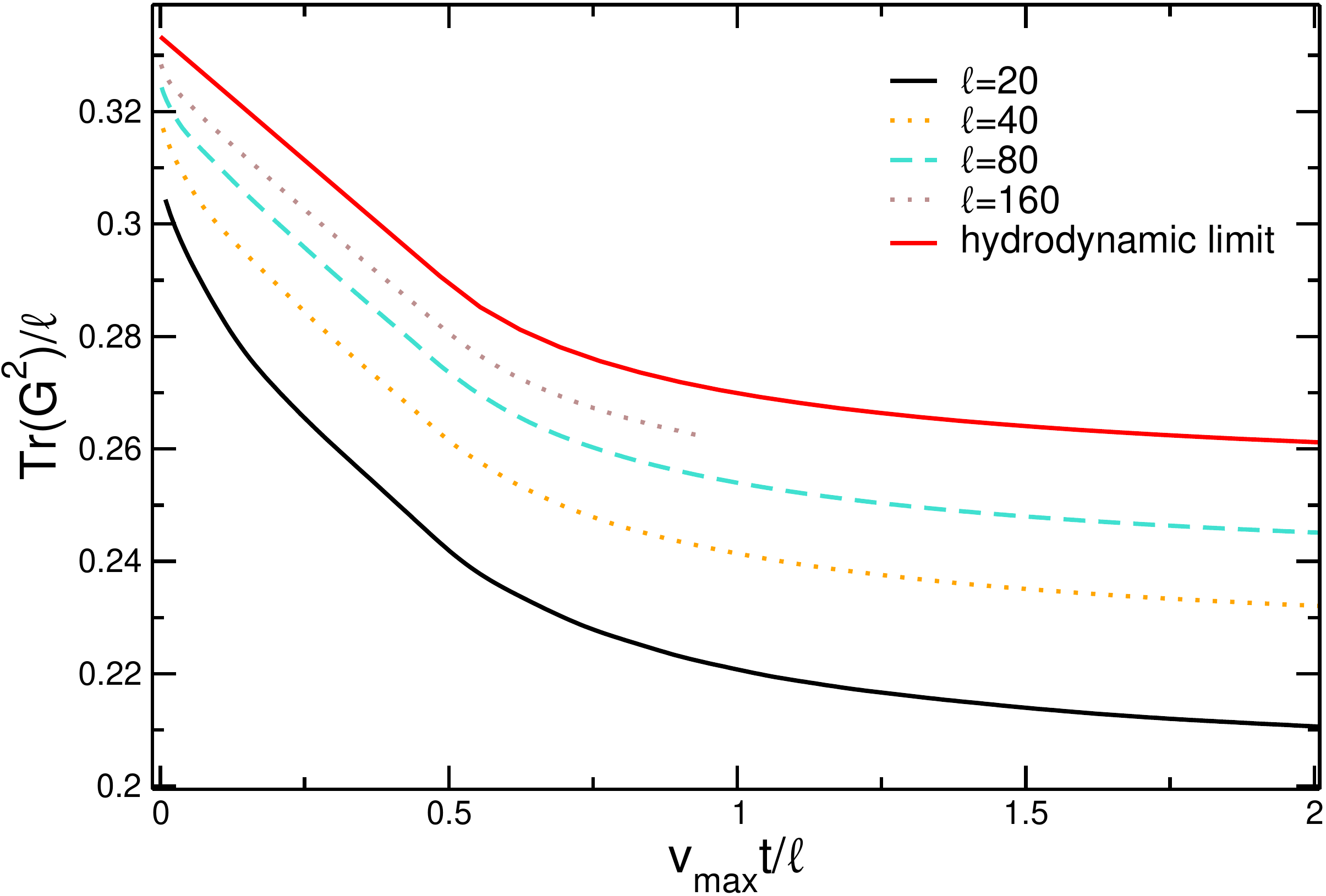}
\caption{ Same as in Fig.~\ref{fig4:traces-check} for the 
	interval $A'$ (centered partition in Fig~\ref{fig0:cartoon} (b)). 
}
\label{fig5:traces-check}
\end{center}
\end{figure}
%########################################
%

Our numerical results for $M_n$ are discussed in Fig.~\ref{fig4:traces-check}. In the panel 
(a) and (b) we plot $M_1$ and $M_2$, respectively. We focus on subsystem $A$ (see Fig.~\ref{fig0:cartoon} (a)). 
The numerical data in the figure are obtained by using~\eqref{eq:G}. We consider the situation in which 
the system is initially prepared in a Fermi sea with $k_F=\pi/3$. Notice that $M_1$ is 
the number of fermions in subsystem $A$. In the absence of dissipation $M_1=\ell k_F/\pi$ 
at any time. As a consequence of the fermion loss the number of particle 
decreases with time. In the figure we report results for several values of 
$\ell$. Clearly, $M_1$ exhibits 
the qualitative behavior discussed in Fig.~\ref{fig1:traces-theo}. At short 
times $t\le \ell/v_{\mathrm{max}}$, $M_1$ decreases linearly, whereas for 
$t\to\infty$ it saturates. However, the data for finite $\ell$ 
exhibit sizeable deviations from the hydrodynamic limit result, which is 
reported as dashed-dotted line in Fig.~\ref{fig4:traces-check}. 
These deviations  are expected. 
The analytic result~\eqref{eq:Mn} is expected to hold only in the hydrodynamic 
limit $t,\ell\to\infty$ with their ratio fixed. Indeed, upon increasing 
$\ell$ the data approach~\eqref{eq:Mn}. 
Importantly, the fact that the initial state is a Fermi sea gives rise 
to logarithmic corrections. This will also happen for the entanglement entropies, 
as we will discuss in section~\ref{sec:num-b}. 
These corrections are visible for $M_2$ (see the inset in 
Fig.~\ref{fig4:traces-check} (b)). In the figure we plot  
the deviation $\delta M_2$ from the hydrodynamic result, which is 
defined as 
\begin{equation}
	\delta M_2:= M_2^\mathrm{hydro}-M_2. 
\end{equation}
We consider the initial deviation at $t=0$. At $t=0$ one expects that 
in the limit $\ell\to\infty$, $M_2=k_f/\pi\ell$. 
The results in the inset of Fig.~\ref{fig4:traces-check} (b) suggest 
the logarithmic behavior as 
\begin{equation}
	\label{eq:fig-1}
	\delta M_2=a_2\ln(\ell)+\dots, 
\end{equation}
with the dots denoting subleading terms, and $a_2$ a constant. 
The dashed-dotted line in the inset of Fig.~\ref{fig4:traces-check} (b) 
is obtained by fitting the constant $a_2$ in~\eqref{eq:fig-1}. The fit 
gives $a_2\approx 0.101$. To our knowledge there is no analytic 
determination of the constant $a_2$, although it should be possible 
by using standard techniques for  free-fermions systems. 
Moreover, although the data in Fig.~\ref{fig4:traces-check} 
(b) suggest that such logarithmic terms survive at finite time, it  is 
not clear a priori whether the constant $a_2$ remains the same. 
Finally, we should remark that the same  logarithmic terms 
should be present for the centered partition in Fig.~\ref{fig0:cartoon} (b). 
Indeed, for $t=0$ the system is translational invariant, and the moments $M_n$ 
do not depend on the position of the subsystem. 
We discuss numerical results for $M_2$ for the centered partition (Fig.~\ref{fig0:cartoon} (b)) 
in Fig.~\ref{fig5:traces-check}. As it is 
clear from the figure, the qualitative behavior is the same as for 
the side bipartition (see Fig.~\ref{fig4:traces-check} (b)). 
Similar finite-size effects as in Fig.~\ref{fig4:traces-check} (b) 
are present. Upon approaching the hydrodynamic limit $t,\ell\to\infty$ deviations 
from the hydrodynamic limit result (red continuous line in the figure) 
vanish. 

%#############################################################
\subsection{Entanglement entropies and logarithmic negativity}
\label{sec:num-b}

%
%########################################
\begin{figure}[t]
\begin{center}
\includegraphics[width=0.5\textwidth]{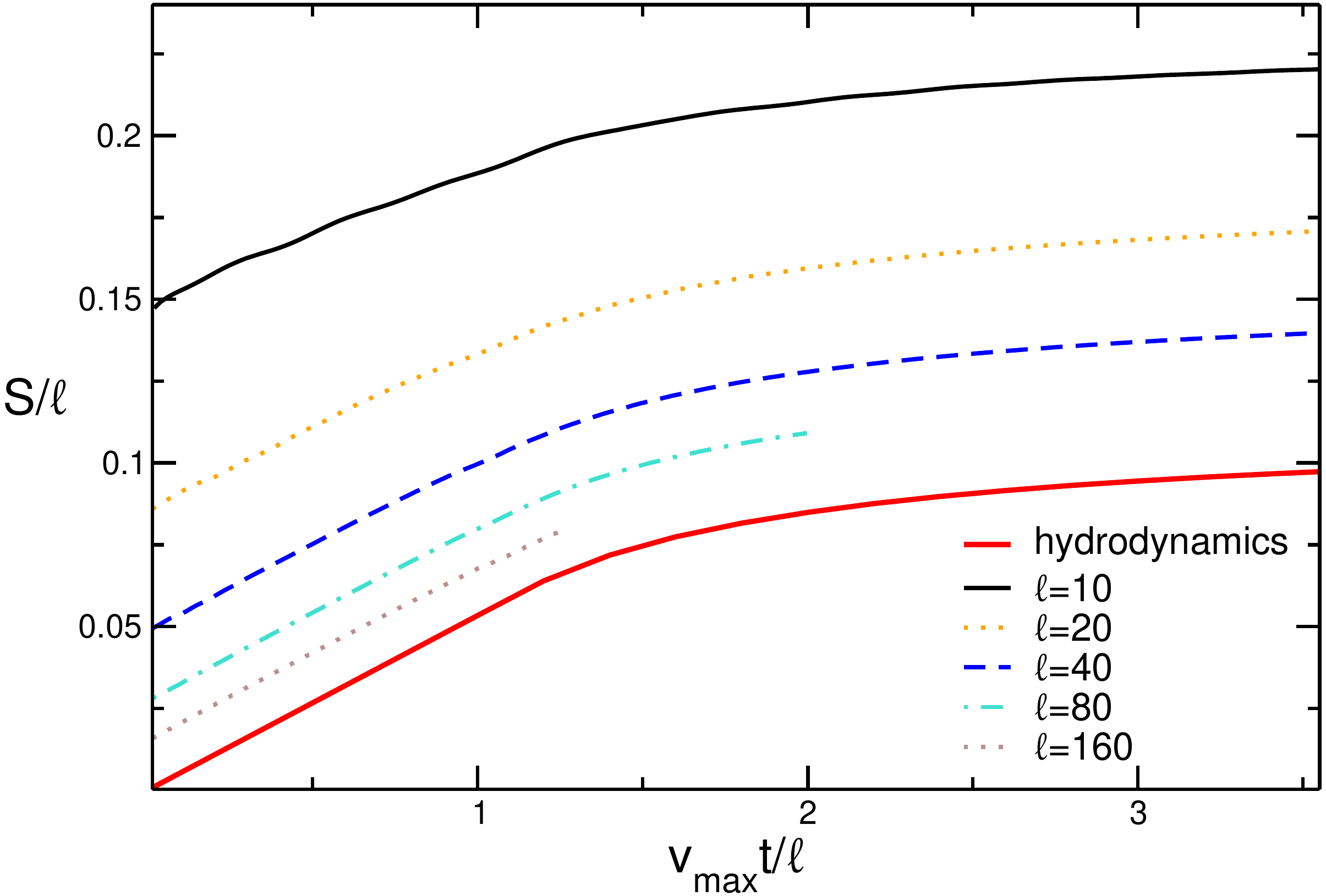}
\caption{ Entanglement entropy $S$ in the fermionic chain 
 subjected to localized losses. We consider  subsystem $A$ 
 (bipartition in Fig.~\ref{fig0:cartoon} (a)). The figure 
 shows the entropy density $S/\ell$ plotted versus $v_\mathrm{max}t/\ell$, 
 with $\ell$ the size of $A$ and $v_\mathrm{max}$ the maximum velocity. 
 All the results are for fixed loss rate $\gamma^-=1$ and $k_F=\pi/3$. 
 We show results for several values of $\ell$, and the analytic 
 result in the hydrodynamic limit (red continuous line in the figure). 
}
\label{fig6:ent}
\end{center}
\end{figure}
%########################################
%

%
%########################################
\begin{figure}[t]
\begin{center}
\includegraphics[width=0.5\textwidth]{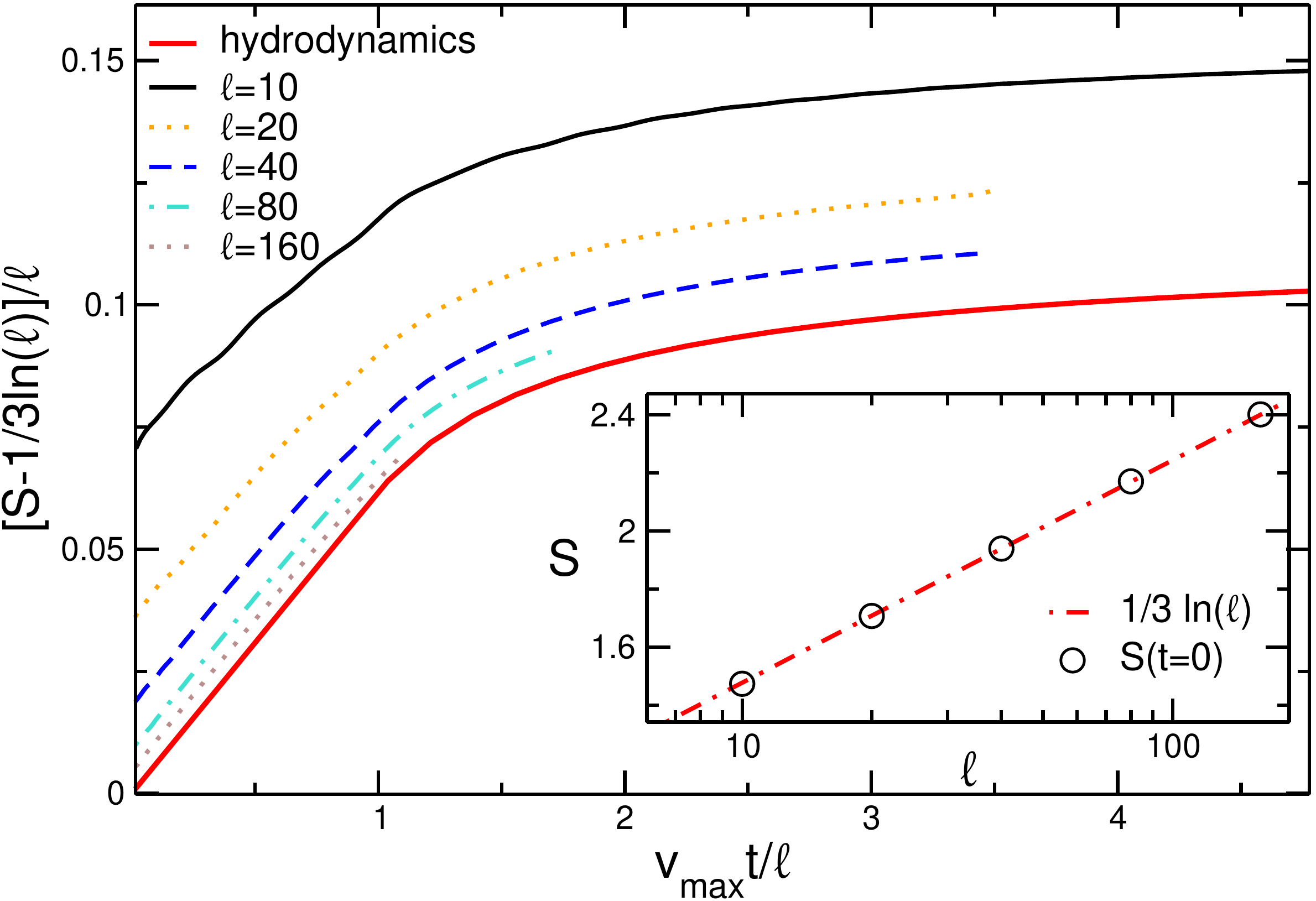}
\caption{ Same data as in Fig.~\ref{fig6:ent} plotting 
 $(S-1/3\ln(\ell))/\ell$, where $1/3\ln(\ell)$ is the 
 initial entanglement entropy. On the $x$-axis 
 $v_\mathrm{max}$ is the maximum velocity and $\ell$ is the size of $A$. Inset: 
 The entanglement entropy at $t=0$ plotted versus 
 $\ell$. Note the logarithmic scale on the $x$-axis. 
 The dashed-dotted line is fit to the CFT prediction 
 $1/3\ln(\ell)+a$, with $a$ a fitting constant. 
}
\label{fig7:ent-subtr}
\end{center}
\end{figure}
%########################################
%

%
%########################################
\begin{figure}[t]
\begin{center}
\includegraphics[width=0.5\textwidth]{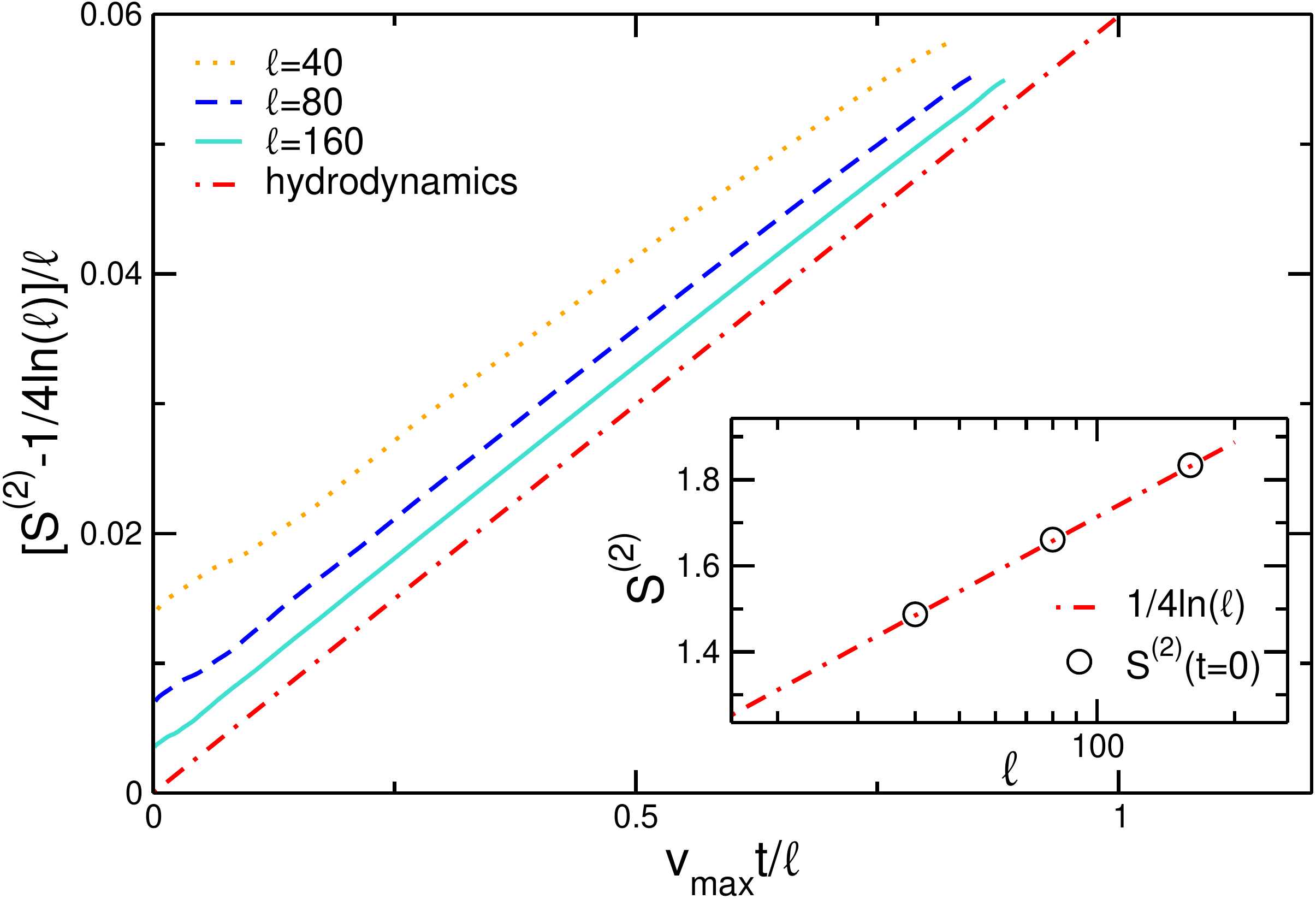}
\caption{ Out-of-equilibrium dynamics of the R\'enyi entropy 
 $S^{(2)}$ of subsystem $A$ (bipartition in Fig.~\ref{fig0:cartoon}) (a). 
 We plot the subtracted entropy $(S-1/4\ln(\ell))/\ell$, with 
 $1/4\ln(\ell)$ the Fermi sea entropy at $t=0$. Data are for 
 $\gamma^-=1$ and $k_F=\pi/3$. On the $x$-axis 
 $v_\mathrm{max}$ is the maximum velocity and $\ell$ is the size of $A$. In the inset we 
 show the entropy at $t=0$ plotted versus $\ell$ to highlight 
 the logarithmic increase. 
}
\label{fig8:renyi}
\end{center}
\end{figure}
%########################################

Let us now discuss the out-of-equilibrium dynamics of the entanglement entropies. 
We first focus on the entanglement entropy for subsystem $A$ next to the dissipation 
source (as in Fig.~\ref{fig0:cartoon} (a)). Our data are reported in Fig.~\ref{fig6:ent}. 
We restrict ourselves to fixed $\gamma^-=1$, plotting the entropy density $S/\ell$ 
versus the rescaled time $v_\mathrm{max}t/\ell$. We show data for $\ell\in[10,160]$. 
We also report the analytic result in the hydrodynamic limit (cf.~\eqref{eq:ent-hydro}). 
Clearly, the numerical data exhibit the expected linear growth for 
$t\le \ell/(v_\mathrm{max})$, followed by a saturation at infinite time. 
Still, one should observe the sizeable deviations from the analytic 
result in the hydrodynamic limit~\eqref{eq:ent-hydro}. This is expected due to the 
finite $\ell$ and finite time $t$. 
Upon approaching the hydrodynamic limit, however, the deviation from~\eqref{eq:ent-hydro} 
decrease. An important remark is that since the initial Fermi sea is a critical state, 
one should expect nontrivial finite-size corrections to the linear entanglement entropy 
growth. For instance, at $t=0$ the entanglement entropies grow logarithmically 
with $\ell$ as in~\eqref{eq:cft-scaling}. 
In Fig.~\ref{fig7:ent-subtr} we subtract the CFT contribution by plotting  
$S-1/3\ln(\ell)$. The data are the same as in Fig.~\ref{fig6:ent}. 
As it is clear from the figure, now the subtracted data exhibit a 
better agreement with the hydrodynamic result. 

We perform a similar analysis for the R\'enyi entropies. In Fig.~\ref{fig8:renyi} we 
show numerical data for the second R\'enyi entropy $S^{(2)}$ plotted versus $v_\mathrm{max}t/\ell$. 
We only consider the bipartition in Fig.~\ref{fig0:cartoon} (a). The data are for 
$\gamma^-=1$ and the initial Fermi sea with $k_F=\pi/3$. As for the von Neumann 
entropy, we subtract the CFT contribution (cf.~\eqref{eq:cft-scaling} with $n=2$) that is present at $t=0$. 
In the Figure we only show data for $v_{\mathrm{max}}t/\ell\lesssim 1$. The agreement with the analytic result 
in the hydrodynamic limit~\eqref{eq:ent-hydro} is satisfactory.

%
%########################################
\begin{figure}[t]
\begin{center}
\includegraphics[width=.9\textwidth]{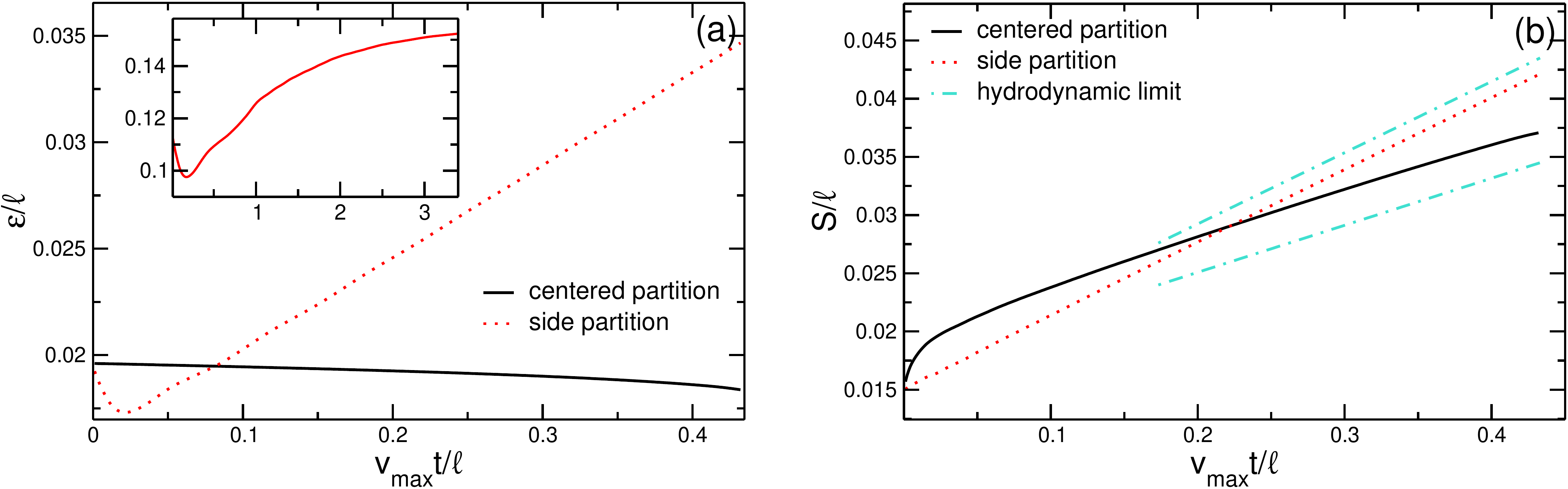}
\caption{ Comparison between the logarithmic negativity ${\cal E}$
 and the entanglement entropy $S$ for the two bipartitions 
 in Fig.~\ref{fig0:cartoon} (a). Panels (a) and (b) 
 show ${\cal E}/\ell$ and $S/\ell$ plotted versus $v_\mathrm{max}t/\ell$, respectively. 
 Here $\ell$ is the size of $A$ and $A'$ (see Fig.~\ref{fig0:cartoon}) and $v_\mathrm{max}$ 
 the maximum velocity. 
 As it is clear from (a) the negativity ${\cal E}$ of $A$ exhibits a linear increase 
 with time, whereas that of $A'$  
 depends mildly on time. Oppositely, the entanglement entropy of both $A$ and 
 $A'$ increases linearly with time (see (b)). 
}
\label{fig9:neg}
\end{center}
\end{figure}
%########################################
%

Finally, it is crucial to compare the dynamics of the entanglement entropy 
with that of the logarithmic negativity (see section~\ref{sec:obs}). As it was stressed 
in section~\ref{sec:obs} the entanglement entropies are not proper entanglement measures in 
the presence of dissipation because the full system is in a mixed state. 
On the other hand, the fermionic negativity ${\cal E}$ (cf.~\eqref{eq:neg-def})
should be sensitive to genuine quantum correlation only. 

As it was anticipated in the introduction, genuine entanglement and statistical correlations are deeply intertwined, 
but it is possible to distinguish them by comparing the behavior of the von Neumann 
entropy and of the logarithmic negativity for the two bipartitions in Fig.~\ref{fig0:cartoon} 
(a) and (b). 
Specifically, subsystem $A$ (see Fig.~\ref{fig0:cartoon} (a)) is 
entangled with its complement because the reflected and the 
transmitted fermions, which form entangled pairs, are shared between 
them. Oppositely, this is not the 
case for $A'$ because the transmitted and the reflected fermions are 
never shared. This scenario implies that the entanglement entropy of 
$A$ and $A'$  exhibit a linear growth with time. On the other hand, 
only the logarithmic negativity of $A$ is expected to grow with time. 

This is demonstrated in Fig.~\ref{fig9:neg} (a) and (b). In Fig.~\ref{fig9:neg} 
(a) we plot the rescaled negativity ${\cal E}/\ell$ versus the rescaled time 
$v_\mathrm{max}t/\ell$, whereas in Fig.~\ref{fig9:neg} (b) we show the 
rescaled entanglement entropy. The data are for fixed $\gamma^-=1$ and 
$k_F=\pi/3$ and subsystem' size $\ell=160$. In both panels we show results 
for the subsystems $A$ (see Fig.~\ref{fig0:cartoon} (a)) 
and $A'$ (see Fig.~\ref{fig0:cartoon} (b)). It is clear from the figure that 
both the negativity and the von Neumann entropy of $A$ increase linearly with time. 
For the von Neumann entropy we report the expected slope of the linear growth 
in the hydrodynamic limit (dashed-dotted line in Fig.~\ref{fig9:neg} (b)), which 
is in perfect agreement with the finite-size numerical results. Notice that at 
asymptotically long times the von Neumann entropy saturates (not shown in the figure), 
as already discussed in the previous sections. 
This saturation happens for the logarithmic 
negativity as well, as expected from the quasiparticle picture discussed above.  
This is shown explicitly in the inset in Fig.~\ref{fig9:neg} (a) for subsystem 
$A$ of length $\ell=20$. As in the main plot we show ${\cal E}/\ell$ versus 
$v_\mathrm{max}t/\ell$.

Let us now discuss the entanglement growth for the bipartition in 
Fig.~\ref{fig0:cartoon} (b). The negativity (see Fig.~\ref{fig9:neg}), 
does not grow with time but it remains almost constant, showing a small 
decreasing trend at long times. Oppositely, 
the entanglement entropy exhibits a linear growth (see Fig.~\ref{fig9:neg} (b)), 
which, again, does not reflect entanglement production. The slope of the linear 
growth (dashed-dotted line) is in agreement with~\eqref{eq:ent-hydro}.

%#############################################################
\section{Conclusions}
\label{sec:concl}

We investigated the interplay between entanglement and statistical correlation in 
a uniform Fermi sea subjected to localized losses. We focused on the hydrodynamic limit 
of long times and the large subsystems, with their ratio fixed. In this 
regime the dynamics of the entanglement entropies can be understood 
analytically. We showed that the logarithmic negativity correctly diagnose the 
production of genuine quantum entanglement, whereas the entanglement 
entropies are sensitive to both quantum as well as classical correlation. 

Let us now illustrate some interesting directions for future research. 
First, our results hold for the Fermi sea as initial state. It should be 
possible to generalize them to other situations, such as finite-temperature 
states, or inhomogeneous initial states, for instance, the domain-wall state. 
One should expect the linear entanglement growth to persist. 
An interesting possibility is to consider the out-of-equilibrium dynamics starting from 
product states. Thus, even in absence of losses the entanglement entropy 
grows linearly with time due to the propagation of entangled pairs 
of quasiparticles. It would be interesting to understand how this 
scenario is modified by localized losses. 
An interesting direction is to try to generalized the hydrodynamic framework to 
the logarithmic negativity, for which it should be possible to obtain a formula similar 
to~\eqref{eq:ent-hydro}. 

Interestingly, our results suggest that local dissipation generically induces 
robust entanglement production.   
An important direction is to try to check this scenario for other types of 
local dissipation. An interesting candidate is 
 incoherent hopping~\cite{eisler2011crossover}. Unlike 
 loss dissipation, for incoherent hopping the Liouvillian describing 
the dynamics of the density matrix is not quadratic. It would interesting 
to understand whether the hydrodynamic approach outlined here 
still applies, at least in the weak dissipation limit. 
Finally, an 
interesting direction would be understand the interplay between entanglement, 
local dissipation, and criticality~\cite{rossini2021coherent}. 

%#############################################################
\appendix
\section{Entanglement entropies in the hydrodynamic limit: Derivation of Eq.~$\eqref{eq:F}$}
\label{sec:app}

In this section we derive formula~\eqref{eq:F}.  
We employ a similar strategy as in Ref.~\cite{calabrese2012quantum}. 
Let us consider the interval $A=[0,\ell]$ (see Fig.~\ref{fig0:cartoon} (a)). 
The main ingredient is the correlation matrix $G_{x,y}$ (cf.~\eqref{eq:f-corr}) 
restricted to $A$, i.e., with $x,y\in A$. 
First, we can rewrite~\eqref{eq:f-corr} as 
\begin{equation}
	\label{eq:Gxy}
	G_{x,y}=\int_{-k_F}^{k_F} \frac{dk}{2\pi} S_{k,x}\bar S_{k,y},
\end{equation}
where we defined 
\begin{equation}
	S_{k,x}=e^{ikx}+r(k)e^{i|kx|}\int_{-\infty}^\infty \frac{dq}{2\pi i} \frac{e^{i(|v_k|t-|x|)q}}{q-i0},
\end{equation}
with $v_k$ the fermion group velocity (cf.~\eqref{eq:v-k}). 
The last term ensures the condition~\eqref{eq:constr} and it 
relies on the well-known identity 
\begin{equation}
	\label{eq:wk}
	\int_{-\infty}^\infty\frac{dq}{2\pi i}\frac{e^{i q x}}{q+i0}=\Theta(x), 
\end{equation}
where $i0$ is a positive convergence factor. 
Let us  define $A_{k,q}$ as 
\begin{equation}
	A_{k,q}(t):=\frac{e^{i t|v_k|q}}{q-i0}r(k). 
\end{equation}
To proceed we use the following identity  
\begin{equation}
	\label{eq:id}
	\sum_{z=1}^\ell e^{izk}=\frac{\ell}{4}\int_{-1}^1d\xi w(k)e^{i(\ell\xi +\ell+1)k/2},\quad 
	\mathrm{with}\,w(k):=\frac{k}{\sin(k/2)}. 
\end{equation}
Let us now define 
\begin{equation}
	\label{eq:F-def}
	F_{k_i,k_j}:=F_{k_i,k_j}^{uu}+F_{k_i,k_j}^{ud}+F_{k_i,k_j}^{du}+F_{k_i,k_j}^{dd}, 
\end{equation}
with 
\begin{align}
	\label{eq:Fuu}
	&F^{uu}_{k_i,k_j}:=\frac{\ell}{4}\int d\xi w(k_i-k_j)e^{i\ell(\xi+1)(k_i-k_j)/2}\\
	&F^{ud}_{k_i,k_j}:=-\frac{\ell}{4}\int d\xi\int\frac{dq}{2\pi i}w(k_i-|k_j|+q) e^{i\ell(k_i-|k_j|+q)(\xi+1)/2}\bar A_{k_j,q}\\
	\label{eq:Fdu}
	&F^{du}_{k_i,k_j}:=\frac{\ell}{4}\int d\xi\int\frac{dq}{2\pi i}w(|k_i|-k_j-q) e^{i\ell(|k_i|-k_j-q)(\xi+1)/2} A_{k_i,q}, 
\end{align}
and 
\begin{multline}
	\label{eq:Fdd}
	F^{dd}_{k_i,k_j}:=-\frac{\ell}{4}\int d\xi\int\frac{dq'}{2\pi i}\int\frac{dq}{2\pi i}
	w(|k_i|-|k_j|-q+q') \\\times e^{i\ell(|k_i|-|k_j|-q+q')(\xi+1)/2} 
	A_{k_i,q}\bar A_{k_j,q}.
\end{multline}
To derive~\eqref{eq:F}, it is convenient to consider the moments of the correlation matrix 
$M_n=\mathrm{Tr}(G^n)$. For generic integer $n$, by using~\eqref{eq:Gxy},~\eqref{eq:F-def} and~\eqref{eq:Fuu}-\eqref{eq:Fdd} 
one obtains that 
\begin{equation}
	\label{eq:f}
	\mathrm{Tr}(G^n)=\int_{-k_F}^{k_F} \frac{d^nk}{(2\pi)^n} \prod_{i=1}^n F_{k_i,k_{i-1}}. 
\end{equation}
Here the variables $k_i$ are arranged in cyclic order, i.e., $k_0=k_n$. 
We can rewrite $F_{k_i,k_j}^{\alpha,\beta}$, with $\alpha,\beta=u,d$ (cf.~\eqref{eq:Fuu}-\eqref{eq:Fdd}) 
as  
\begin{align}
	\label{eq:fuu}
	&F^{uu}_{k_i,k_j}=\frac{\ell}{2}\int d\xi e^{i\ell(\xi+1)(k_i-k_j)/2}\\
	&F^{ud}_{k_i,k_j}=\frac{\ell}{2}\int d\xi e^{i\ell(k_i-|k_j|)(\xi+1)/2}r(k_j)\Theta(-\ell(\xi+1)/2+|v_{k_j}|t)\\
	&F^{du}_{k_i,k_j}=\frac{\ell}{2}\int d\xi e^{i\ell(|k_i|-k_j)(\xi+1)/2} r(k_i)\Theta(-\ell(\xi+1)/2+|v_{k_i}|t), 
\end{align}
and 
\begin{multline}
\label{eq:fdd}
	F^{dd}_{k_i,k_j}=\frac{\ell}{2}\int d\xi e^{i\ell(|k_i|-|k_j|)(\xi+1)/2} r(k_i)r(k_j)\\\times
	\Theta(-\ell(\xi+1)/2+|v_{k_i}|t)\Theta(-\ell(\xi+1)/2+|v_{k_j}|t). 
\end{multline}
To obtain~\eqref{eq:fuu}-\eqref{eq:fdd}, we used that in the hydrodynamic 
limit $\ell,t\to\infty$ with the ratio $t/\ell$ fixed the integrals in~\eqref{eq:f} 
are dominated by the regions with  $k_i\to k_j$ and $q\to0$. 
This implies that $w(k_i-k_j)\to 1/2$ (cf.~\eqref{eq:id}), and that one can perform the integration over 
$q$ and $q'$ in~\eqref{eq:Fuu}-\eqref{eq:Fdd}, which, by using~\eqref{eq:wk}, give the Heaviside 
theta functions in~\eqref{eq:fuu}-\eqref{eq:fdd}. 
We can now rewrite~\eqref{eq:f} as 
\begin{equation}
	\label{eq:compl}
	\mathrm{Tr}(G^n)=\Big(\frac{\ell}{2}\Big)^n\int_{-k_F}^{k_F}\frac{d^n k}{(2\pi)^n}\int_{-1}^1 d^n\xi\prod_{i=1}^n\widetilde F_{k_i,k_{i-1}}(\xi_i). 
\end{equation}
Here we defined $\widetilde F_{k_i,k_{i-1}}:=\widetilde F_{k_i,k_j}^{uu}+
\widetilde F_{k_i,k_j}^{ud}+\widetilde F_{k_i,k_j}^{du}+\widetilde F_{k_i,k_j}^{dd}$, 
where $\widetilde F_{k_i,k_j}^{\alpha,\beta}$ with $\alpha,\beta=u,d$ are 
the integrands appearing in~\eqref{eq:fuu}-\eqref{eq:fdd}. 
To proceed, we now treat the integrals over $\xi_i$ by using the stationary phase 
approximation in the hydrodynamic limit. 
We first observe that~\eqref{eq:compl} can be rewritten as 
\begin{multline}
	\label{eq:compl-1}
	\mathrm{Tr}(G^n)=\Big(\frac{\ell}{2}\Big)^n\int_{-k_F}^{k_F}\frac{d^n k}{(2\pi)^n}\int_{-1}^1 d^n\xi
	\\\sum\limits_{\sigma_i,\tau_i=0,1}\,\prod_{i=1}^n e^{i\ell (\xi_i+1)(k_{\sigma_i}-k_{\tau_{i-1}})/2} \tilde 
	r^{\sigma_i}(\xi_i,k_{\sigma_i})\tilde r^{\tau_{i-1}}(\xi_i,k_{\tau_{i-1}}). 
\end{multline}
Here we defined 
\begin{equation}
	\tilde r(\xi,k)= r(k)\Theta(-\ell(\xi+1)/2+|v_{k}|t). 
\end{equation}
Here we also defined 
\begin{equation}
	\label{eq:ksigma}
	k_{\sigma_i}=\left\{\begin{array}{cc}
			k_i   &\mathrm{if}\,\sigma_i=0\\
			|k_i| &\mathrm{if}\,\sigma_i=1
	\end{array}\right.
\end{equation}
The same definition as~\eqref{eq:ksigma} holds for $k_{\tau_i}$. 
Eq.~\eqref{eq:compl-1} arises directly from~\eqref{eq:compl}. Each term $\widetilde 
F_{k_i,k_j}$ in~\eqref{eq:compl} contains a phase factor $e^{i\ell(\xi_i+1)(k_i-k_{i-1})}$, 
where $k_i$, $k_{i-1}$ can be replaced by $|k_i|,|k_{i-1}|$. Each term $|k_i|$ is 
accompanied by a factor $\tilde r(\xi_i,k_i)$. The sum over $\sigma_i,\tau_i$ in~\eqref{eq:compl-1} 
accounts for all the possible ways of distributing the terms with the absolute values 
$|k_i|$. 

We first focus on the situation with $\sigma_i=\tau_i=0$ for any $i$. 
Thus, within the stationary phase approximation the integral in~\eqref{eq:compl-1} 
in the large $\ell$ limit is dominated by the stationary points of the 
exponent of the phase factor. By imposing stationarity with respect to the 
variables $\xi_i$, one obtains that 
\begin{equation}
	k_i=k_1,\quad\forall i. 
\end{equation}
Now~\eqref{eq:compl-1} becomes 
\begin{equation}
\label{eq:compl-0}
{\mathcal I}^{(n)}_{0}=\Big(\frac{\ell}{2}\Big)^n\int_{-k_F}^{k_F}\frac{d^nk}{(2\pi)^n}\int_{-1}^1 d^n\xi
	e^{i\ell\sum_{i=1}^n(\xi_i+1)(k_i-k_{i-1})/2}. 
\end{equation}
Although the integral~\eqref{eq:compl-0} can be computed exactly, 
it is useful to discuss the stationary phase approximation. 
Let us change variables as 
\begin{align}
&\zeta_1:=\xi_1\\
&\zeta_i:=\xi_{i+1}-\xi_i,\quad i\in[1,n]. 
\end{align}
The variables $\xi_i$ and $\zeta_i$ satisfy cyclic boundary conditions. 
We obtain that~\eqref{eq:compl-0} is rewritten as 
\begin{equation}
	\label{eq:integ}
	{\mathcal I}^{(n)}_0=	\Big(\frac{\ell}{2}\Big)^n\int_{-k_F}^{k_F}\frac{d^nk}{(2\pi)^n}\int d^n\zeta
	e^{-i\ell\sum_{j=1}^{n}\zeta_j(k_j-k_{1})/2}. 
\end{equation}
Notice that  the $1$ in $(\xi_i+1)$ in~\eqref{eq:compl-0} cancels out in the 
sum over $i$, and it would also be irrelevant at the stationary 
point where $k_i\to k_1$ for any $i$. The integrand 
in~\eqref{eq:integ} does not depend on $\zeta_1$. 
The integration domain for the variables $\zeta_1$ is given as 
\begin{equation}
	-1\le \zeta_1-\sum_{j=k}^n\zeta_j \le 1,\quad \forall k\in[2,n]. 
\end{equation}
As the integrand in~\eqref{eq:integ} does not depend on $\zeta_1$, one can 
perform the integral to obtain 
\begin{multline}
	\label{eq:integ-1}
	{\mathcal I}_0^{(n)}:=\Big(\frac{\ell}{2}\Big)\int_{-k_F}^{k_F}\frac{dk_1}{2\pi}\Lambda_0^{(n-1)}(k_1)=\\
	\Big(\frac{\ell}{2}\Big)^n\int_{-k_F}^{k_F}\frac{dk_1}{2\pi}
	\int_{-k_F}^{k_F}\frac{d^{n-1}k}{(2\pi)^{n-1}}\int d^{n-1}\zeta
	e^{-i\ell\sum_{j=1}^{n}\zeta_j(k_j-k_{1})/2}\mu(\{\zeta_k\}), 
\end{multline}
where we also isolated the integration over $k_1$. Here $\mu(\{\zeta_k\})$ is the measure 
of the allowed values for $\zeta_1$, and it reads as 
\begin{equation}
	\mu(\{\zeta_k\})=\max\Big[0,\min\limits_{k\in[2,n]}\Big[1-\sum_{j=k}^n\zeta_j\Big]+
	\min\limits_{k\in[2,n]}\Big[1+\sum_{j=k}^n\zeta_j\Big]\Big]. 
\end{equation}
We can now apply the stationary phase approximation to the integral 
\begin{equation}
	\Lambda^{(n-1)}_0=\Big(\frac{\ell}{2}\Big)^{n-1}\int_{-k_F}^{k_F}\frac{d^{n-1}k}{(2\pi)^{n-1}}\int d^{n-1}\zeta
	e^{-i\ell\sum_{j=1}^{n}\zeta_j(k_j-k_{1})/2}\mu(\{\zeta_k\}). 
\end{equation}
The stationary phase approximation states that for sufficiently smooth $N$-dimensional 
functions $f(\vec x)$ and $g(\vec x)$, in the limit $\ell\to\infty$  one has~\cite{wong} 
\begin{equation}
	\label{eq:s-phase}
	\int_\Omega d^Nx g(\vec x)e^{i\ell f(\vec x)}=\Big(\frac{2\pi}{\ell}\Big)^{N/2}
	g(\vec x_0)|\det H|^{-1/2}e^{i\ell f(\vec x_0)+\frac{i\pi\sigma}{4}},  
\end{equation}
where $\Omega$ is the integration domain, $\vec x_0$  is the stationary 
point of $f(\vec x)$, i.e., such that $\nabla f(\vec x)=0$, 
$H$ is the Hessian matrix, and $\sigma$ its signature, i.e., the difference 
between the number of positive and negative eigenvalues. 
A straightforward application of the stationary phase gives that in the 
limit $\ell\to\infty$ the $\Lambda_0^{(n-1)}$ is dominated by the stationary 
point as 
\begin{align}
	\label{eq:k-sp}
& \bar k_j=k_1,\quad j=2,\dots,n,\\
\label{eq:z-p}
& \bar \zeta_j=0,\quad j=2,\dots,n. 
\end{align}
In our case the phase in~\eqref{eq:s-phase} vanishes and the signature of the Hessian 
is zero. Moreover, $\det H=2^{-2n+2}$.  Putting everything together we obtain that 
\begin{equation}
	\label{eq:L0}
	\Lambda^{(n-1)}_0=2, 
\end{equation}
where we used that $\mu(\{\zeta_k\})=2$ at the stationary point. 
Note that there is no dependence on $k_1$ in~\eqref{eq:L0}. 
Finally, we obtain  that the integral~\eqref{eq:integ} is given as 
\begin{equation}
	{\mathcal I}_0^{(n)}=\ell\int_{-k_F}^{k_F}\frac{dk_1}{2\pi}. 
\end{equation}
Let us now consider the generic integral~\eqref{eq:compl-1}.  
We now observe that for any pair of indices $(\sigma_i,\tau_i)$ 
there are two possible situations that can occur. Specifically, 
we define $(\sigma_i,\tau_i)$ as {\it paired} if $\sigma_i=\tau_i=1$, 
whereas we define $(\sigma_i,\tau_i)$ as {\it unpaired} otherwise. 
Notice that if $(\sigma_i,\tau_i)$ are paired it means that both occurrences 
of momentum $k_i$ appear with the absolute value $|k_i|$ in~\eqref{eq:compl-1}. 

It is straightforward to convince oneself that the presence of  a 
single set of unpaired indices $(\sigma_i,\tau_i)$ implies that 
in the limit $\ell\to\infty$ the stationary point is given as 
\begin{equation}
	\label{eq:statio-2}
k_i=k_1>0,\quad\forall i,  
\end{equation}
i.e., all the momenta have to be positive to have a finite contribution 
in the stationary phase. This implies that one can remove the absolute 
values of the momenta. Then, the derivation of the stationary phase is 
similar to that for the case with $\sigma_i=\tau_i=0,\forall i$.  

An important difference is that for any index $\sigma_i=1$ and 
$\tau_i=1$ there is a factor $\tilde r(\xi_i,k_i)$. This implies that 
the integration over $\zeta_1$ in principle cannot be performed as in~\eqref{eq:L0}. 
However, at the stationary point, from~\eqref{eq:z-p} one obtains that 
$\xi_i\to\xi_1=\zeta_1$ and $k_i\to k_1$ for any $i$. One is left with the integral 
over $\zeta_1$ as 
\begin{equation}
	\label{eq:inte}
	\int_{-1}^1d\zeta_1\Theta(-\ell(\zeta_1+1)/2+t|v_k|)=
	2\min(|v_k|t/\ell,1),\quad\textrm{with}\, k>0.  
\end{equation}
Let us now discuss what happens when paired indices are present. 
It is clear that the main consequence of the presence of 
paired indices $\sigma_i,\tau_i$ is a factor $2$ because the 
integrands do not depend on the sign of the momenta. 
To discuss the result of the stationary phase, let us define for the following 
the number of paired momenta as $p$, and the total number of 
momenta appearing with the absolute value as $n_a$. Let us consider 
the case $n_a>0$ since the case with $n_a=0$ was treated above.  
$n_a$ is the number of $\sigma_i=1$ and $\tau_i=1$.  
The number of terms $N_{(n,n_a,p)}$ with fixed $n,n_a,p$ can be obtained by 
elementary combinatorics as 
\begin{equation}
\label{eq:f1}
N_{(n,n_a,p)}=\binom{n}{p}\binom{n-p}{n_a-2p}2^{n_a-2p}. 
\end{equation}
Now we use take into account that for each set of paired indices 
there is a factor two. By summing over the possible number of pairs $p$, 
we obtain the total number of terms $N'_{(n,n_a)}$ as 
\begin{equation}
\label{eq:f2}
N_{(n,n_a)}'=\sum\limits_{p=0}^{\lfloor n_a/2\rfloor} 2^{p}N_{(n,n_a,p)}. 
\end{equation}
We now use that for each $n_a$ there is term $r^{n_a}$. 
Finally, it is straightforward to perform the sum over $n_a$ to obtain the 
total contribution as 
\begin{equation}
	\mathrm{Tr}(G^n)=\Big(\frac{\ell}{2}\Big)\int_{-k_F}^{k_F}\frac{dk_1}{2\pi}
	(\Lambda^{(n-1)}_0+\Lambda^{(n-1)})
\end{equation}
where $\Lambda_0^{(n-1)}=2$ and 
\begin{equation}
	\label{eq:lambda}
	\Lambda^{(n-1)}=
	2\min(|v_{k_1}|t/\ell,1)\Big(\sum\limits_{n_a=1}^{2n} N'_{(n,n_a)}r^{n_a}-1\Big)\Theta(k_1). 
\end{equation}
We now use that 
\begin{equation}
\label{eq:f3}
\sum\limits_{n_a=1}^{2n} N'_{(n,n_a)}r^{n_a}=(1+2r+2r^2)^n-1. 
\end{equation}
where $r(k_1)$ is the reflection amplitude in~\eqref{eq:delta-coeff}. 
We can also use that 
\begin{equation}
	1+2r+2r^2=1-|a|^2,  
\end{equation}
where $|a(k)|^2$ is the absorption coefficient. 
Thus, putting everything together one obtains the final formula 
for $\mathrm{Tr}(G^n_A)$ as 
\begin{equation}
	\label{eq:tr-res}
	\mathrm{Tr}(G^n_A)=
	\ell\int_{-k_F}^{k_F}\frac{dk}{2\pi}\Big[1-\frac{1}{2}\min(|v_k|t/\ell,1)+\frac{1}{2}(1-|a(k)|^2)^n\min(|v_k|t/\ell,1)\Big]. 
\end{equation}
Here we replaced $k_1\to k$ and we used the fact that the integrand is symmetric 
under $k\to-k$ to remove the factor $\Theta(k)$ in~\eqref{eq:lambda}. 
The subscript $A$ in~\eqref{eq:tr-res} is to stress that it holds for subsystem 
$A$ (see Fig.~\ref{fig0:cartoon} (a)). 
Crucially, Eq.~\eqref{eq:tr-res} depends only on the local density of 
fermions $1-|a|^2$ that are not absorbed at the origin. 

Finally, we comment on the modifications in order to generalize~\eqref{eq:tr-res} to 
the case of the bipartition in Fig.~\ref{fig0:cartoon} (b), i.e., for the interval 
$A'$ centered around the impurity. 
The main difference is that~\eqref{eq:f2} has to be replaced by 
$\widetilde N'_{(n,n_a)}$ as 
\begin{equation}
	\widetilde N'_{(n,n_a)}=\frac{2^{n_a}}{4}\binom{2n}{n_a}. 
\end{equation}
A straightforward generalization of the steps leading to~\eqref{eq:tr-res} gives  
\begin{equation}
	\label{eq:tr-1-res}
	\mathrm{Tr}(G^n_{A'})=\ell\int_{-k_F}^{k_F}
	\frac{dk}{2\pi}\Big[1-\frac{1}{4}\min(2|v_k|t/\ell,1)+\frac{1}{4}(1-2|a(k)|^2)^n\min(2|v_k|t/\ell,1)\Big]. 
\end{equation}
The factor $1/4$ and the $2|v_k|$ in the integrand in~\eqref{eq:tr-1-res} reflect 
that the distance from the impurity and the edges of subsystem $A'$ is $\ell/2$ and 
not $\ell$ as in~\eqref{eq:tr-res}. Furthermore, the factor $1-2|a|^2$ instead 
of $1-|a|^2$ in~\eqref{eq:tr-res} arises because fermions are absorbed from 
both sides of the impurity. 

Finally, by useing~\eqref{eq:tr-res} and~\eqref{eq:tr-1-res} one can 
obtain the hydrodynamic behavior  of 
\begin{equation}
	\mathrm{Tr}({\mathcal F}(G))
\end{equation}
where ${\mathcal F}(z)$ is smooth enough to admit a Taylor expansion around 
$z=0$. By Taylor expanding ${\mathcal F}(z)$ for $z=0$ and by using~\eqref{eq:tr-res} 
and~\eqref{eq:tr-1-res}, one obtains~\eqref{eq:F}.  

\bibliographystyle{SciPost_bibstyle.bst}
\bibliography{bibliography}

\begin{thebibliography}{10}
\providecommand{\url}[1]{\texttt{#1}}
\providecommand{\urlprefix}{URL }
\expandafter\ifx\csname urlstyle\endcsname\relax
  \providecommand{\doi}[1]{doi:\discretionary{}{}{}#1}\else
  \providecommand{\doi}{doi:\discretionary{}{}{}\begingroup
  \urlstyle{rm}\Url}\fi
\providecommand{\eprint}[2][]{\url{#2}}

\bibitem{lin2013}
Y.~Lin, J.~P. Gaebler, F.~Reiter, T.~R. Tan, R.~Bowler, A.~S. S{\o}rensen,
  D.~Leibfried and D.~J. Wineland,
\newblock \emph{Dissipative production of a maximally entangled steady state of
  two quantum bits},
\newblock Nature \textbf{504}(7480), 415 (2013),
\newblock \doi{10.1038/nature12801}.

\bibitem{verstraete-2009}
F.~Verstraete, M.~M. Wolf and J.~Ignacio~Cirac,
\newblock \emph{Quantum computation and quantum-state engineering driven by
  dissipation},
\newblock Nature Physics \textbf{5}(9), 633 (2009),
\newblock \doi{10.1038/nphys1342}.

\bibitem{diehl-2011}
S.~Diehl, E.~Rico, M.~A. Baranov and P.~Zoller,
\newblock \emph{Topology by dissipation in atomic quantum wires},
\newblock Nature Physics \textbf{7}(12), 971 (2011),
\newblock \doi{10.1038/nphys2106}.

\bibitem{preskill2018quantum}
J.~Preskill,
\newblock \emph{Quantum {C}omputing in the {NISQ} era and beyond},
\newblock {Quantum} \textbf{2}, 79 (2018),
\newblock \doi{10.22331/q-2018-08-06-79}.

\bibitem{petruccione}
H.~P. Breuer and F.~Petruccione,
\newblock \emph{The theory of open quantum systems},
\newblock Oxford University Press, Great Clarendon Street (2002).

\bibitem{prosen-2008}
T.~Prosen,
\newblock \emph{{Third quantization: a general method to solve master equations
  for quadratic open Fermi systems}},
\newblock New Journal of Physics \textbf{10}(4), 43026 (2008),
\newblock \doi{10.1088/1367-2630/10/4/043026}.

\bibitem{prosen-2011}
T.~c.~v. Prosen,
\newblock \emph{Exact nonequilibrium steady state of a strongly driven open
  $xxz$ chain},
\newblock Phys. Rev. Lett. \textbf{107}, 137201 (2011),
\newblock \doi{10.1103/PhysRevLett.107.137201}.

\bibitem{prosen-2014}
T.~c.~v. Prosen,
\newblock \emph{Exact nonequilibrium steady state of an open hubbard chain},
\newblock Phys. Rev. Lett. \textbf{112}, 030603 (2014),
\newblock \doi{10.1103/PhysRevLett.112.030603}.

\bibitem{prosen-2015}
T.~Prosen,
\newblock \emph{{Matrix product solutions of boundary driven quantum chains}},
\newblock Journal of Physics A: Mathematical and Theoretical \textbf{48}(37),
  373001 (2015),
\newblock \doi{10.1088/1751-8113/48/37/373001}.

\bibitem{znidaric-2010}
M.~{\v{Z}}nidari{\v{c}},
\newblock \emph{{Exact solution for a diffusive nonequilibrium steady state of
  an open quantum chain}},
\newblock Journal of Statistical Mechanics: Theory and Experiment
  \textbf{2010}(05), L05002 (2010),
\newblock \doi{10.1088/1742-5468/2010/05/l05002}.

\bibitem{znidaric-2011}
M.~\ifmmode \check{Z}\else \v{Z}\fi{}nidari\ifmmode~\check{c}\else \v{c}\fi{},
\newblock \emph{Solvable quantum nonequilibrium model exhibiting a phase
  transition and a matrix product representation},
\newblock Phys. Rev. E \textbf{83}, 011108 (2011),
\newblock \doi{10.1103/PhysRevE.83.011108}.

\bibitem{medvedyeva-2016}
M.~V. Medvedyeva, F.~H.~L. Essler and T.~c.~v. Prosen,
\newblock \emph{Exact bethe ansatz spectrum of a tight-binding chain with
  dephasing noise},
\newblock Phys. Rev. Lett. \textbf{117}, 137202 (2016),
\newblock \doi{10.1103/PhysRevLett.117.137202}.

\bibitem{buca-2020}
B.~Buca, C.~Booker, M.~Medenjak and D.~Jaksch,
\newblock \emph{Dissipative bethe ansatz: Exact solutions of quantum many-body
  dynamics under loss} (2020), \eprint{2004.05955}.

\bibitem{bastianello-2020}
A.~Bastianello, J.~De~Nardis and A.~De~Luca,
\newblock \emph{Generalized hydrodynamics with dephasing noise},
\newblock Phys. Rev. B \textbf{102}, 161110 (2020),
\newblock \doi{10.1103/PhysRevB.102.161110}.

\bibitem{essler-2020}
F.~H.~L. Essler and L.~Piroli,
\newblock \emph{Integrability of one-dimensional lindbladians from
  operator-space fragmentation},
\newblock Phys. Rev. E \textbf{102}, 062210 (2020),
\newblock \doi{10.1103/PhysRevE.102.062210}.

\bibitem{ziolkowska-2020}
A.~A. Ziolkowska and F.~H. Essler,
\newblock \emph{{Yang-Baxter integrable Lindblad equations}},
\newblock SciPost Phys. \textbf{8}, 44 (2020),
\newblock \doi{10.21468/SciPostPhys.8.3.044}.

\bibitem{sieberer-2016}
L.~M. Sieberer, M.~Buchhold and S.~Diehl,
\newblock \emph{{Keldysh field theory for driven open quantum systems}},
\newblock Reports on Progress in Physics \textbf{79}(9), 96001 (2016),
\newblock \doi{10.1088/0034-4885/79/9/096001}.

\bibitem{bouchoule-2020}
I.~Bouchoule, B.~Doyon and J.~Dubail,
\newblock \emph{{The effect of atom losses on the distribution of rapidities in
  the one-dimensional Bose gas}},
\newblock SciPost Phys. \textbf{9}, 44 (2020),
\newblock \doi{10.21468/SciPostPhys.9.4.044}.

\bibitem{Friedman_2020}
A.~J. Friedman, S.~Gopalakrishnan and R.~Vasseur,
\newblock \emph{Diffusive hydrodynamics from integrability breaking},
\newblock Physical Review B \textbf{101}(18) (2020).

\bibitem{deleeuw-2021}
M.~de~Leeuw, C.~Paletta and B.~Pozsgay,
\newblock \emph{Constructing integrable lindblad superoperators} (2021),
  \eprint{2101.08279}.

\bibitem{denardis2021}
J.~D. Nardis, S.~Gopalakrishnan, R.~Vasseur and B.~Ware,
\newblock \emph{Stability of superdiffusion in nearly integrable spin chains}
  (2021), \eprint{2102.02219}.

\bibitem{bertini-2016}
B.~Bertini, M.~Collura, J.~De~Nardis and M.~Fagotti,
\newblock \emph{Transport in out-of-equilibrium $xxz$ chains: Exact profiles of
  charges and currents},
\newblock Phys. Rev. Lett. \textbf{117}, 207201 (2016),
\newblock \doi{10.1103/PhysRevLett.117.207201}.

\bibitem{olalla-2016}
O.~A. Castro-Alvaredo, B.~Doyon and T.~Yoshimura,
\newblock \emph{Emergent hydrodynamics in integrable quantum systems out of
  equilibrium},
\newblock Phys. Rev. X \textbf{6}, 041065 (2016),
\newblock \doi{10.1103/PhysRevX.6.041065}.

\bibitem{alba-2021}
V.~Alba and F.~Carollo,
\newblock \emph{Spreading of correlations in markovian open quantum systems},
\newblock Phys. Rev. B \textbf{103}, L020302 (2021),
\newblock \doi{10.1103/PhysRevB.103.L020302}.

\bibitem{maity-2020}
S.~Maity, S.~Bandyopadhyay, S.~Bhattacharjee and A.~Dutta,
\newblock \emph{Growth of mutual information in a quenched one-dimensional open
  quantum many-body system},
\newblock Phys. Rev. B \textbf{101}, 180301 (2020),
\newblock \doi{10.1103/PhysRevB.101.180301}.

\bibitem{calabrese-2005}
P.~Calabrese and J.~Cardy,
\newblock \emph{{Evolution of entanglement entropy in one-dimensional
  systems}},
\newblock Journal of Statistical Mechanics: Theory and Experiment
  \textbf{2005}(04), P04010 (2005),
\newblock \doi{10.1088/1742-5468/2005/04/p04010}.

\bibitem{fagotti-2008}
M.~Fagotti and P.~Calabrese,
\newblock \emph{Evolution of entanglement entropy following a quantum quench:
  Analytic results for the $xy$ chain in a transverse magnetic field},
\newblock Phys. Rev. A \textbf{78}, 010306 (2008),
\newblock \doi{10.1103/PhysRevA.78.010306}.

\bibitem{alba-2017}
V.~Alba and P.~Calabrese,
\newblock \emph{{Entanglement and thermodynamics after a quantum quench in
  integrable systems}},
\newblock Proceedings of the National Academy of Sciences of the United States
  of America \textbf{114}(30) (2017),
\newblock \doi{10.1073/pnas.1703516114}.

\bibitem{alba-2018}
V.~Alba and P.~Calabrese,
\newblock \emph{{Entanglement dynamics after quantum quenches in generic
  integrable systems}},
\newblock SciPost Phys. \textbf{4}, 17 (2018),
\newblock \doi{10.21468/SciPostPhys.4.3.017}.

\bibitem{alba2017quench}
V.~Alba and P.~Calabrese,
\newblock \emph{Quench action and r\'enyi entropies in integrable systems},
\newblock Phys. Rev. B \textbf{96}, 115421 (2017),
\newblock \doi{10.1103/PhysRevB.96.115421}.

\bibitem{alba2017renyi}
V.~Alba and P.~Calabrese,
\newblock \emph{R{\'e}nyi entropies after releasing the n{\'e}el state in the
  xxz spin-chain},
\newblock Journal of Statistical Mechanics: Theory and Experiment
  \textbf{2017}(11), 113105 (2017).

\bibitem{alba2018entanglement}
V.~Alba,
\newblock \emph{Entanglement and quantum transport in integrable systems},
\newblock Physical Review B \textbf{97}(24), 245135 (2018).

\bibitem{alba2021generalizedhydrodynamic}
V.~Alba, B.~Bertini, M.~Fagotti, L.~Piroli and P.~Ruggiero,
\newblock \emph{Generalized-hydrodynamic approach to inhomogeneous quenches:
  Correlations, entanglement and quantum effects} (2021), \eprint{2104.00656}.

\bibitem{dolgirev-2020}
P.~E. Dolgirev, J.~Marino, D.~Sels and E.~Demler,
\newblock \emph{Non-gaussian correlations imprinted by local dephasing in
  fermionic wires},
\newblock Phys. Rev. B \textbf{102}, 100301 (2020),
\newblock \doi{10.1103/PhysRevB.102.100301}.

\bibitem{jin-2020}
T.~Jin, M.~Filippone and T.~Giamarchi,
\newblock \emph{Generic transport formula for a system driven by markovian
  reservoirs},
\newblock Phys. Rev. B \textbf{102}, 205131 (2020),
\newblock \doi{10.1103/PhysRevB.102.205131}.

\bibitem{maimbourg-2020}
T.~Maimbourg, D.~M. Basko, M.~Holzmann and A.~Rosso,
\newblock \emph{Bath-induced zeno localization in driven many-body quantum
  systems} (2020), \eprint{2009.11784}.

\bibitem{froml-2019}
H.~Fr\"oml, A.~Chiocchetta, C.~Kollath and S.~Diehl,
\newblock \emph{Fluctuation-induced quantum zeno effect},
\newblock Phys. Rev. Lett. \textbf{122}, 040402 (2019),
\newblock \doi{10.1103/PhysRevLett.122.040402}.

\bibitem{tonielli-2019}
F.~Tonielli, R.~Fazio, S.~Diehl and J.~Marino,
\newblock \emph{Orthogonality catastrophe in dissipative quantum many-body
  systems},
\newblock Phys. Rev. Lett. \textbf{122}, 040604 (2019),
\newblock \doi{10.1103/PhysRevLett.122.040604}.

\bibitem{froml-2020}
H.~Fr\"oml, C.~Muckel, C.~Kollath, A.~Chiocchetta and S.~Diehl,
\newblock \emph{Ultracold quantum wires with localized losses: Many-body
  quantum zeno effect},
\newblock Phys. Rev. B \textbf{101}, 144301 (2020),
\newblock \doi{10.1103/PhysRevB.101.144301}.

\bibitem{krapivsky-2019}
P.~L. Krapivsky, K.~Mallick and D.~Sels,
\newblock \emph{Free fermions with a localized source},
\newblock Journal of Statistical Mechanics: Theory and Experiment
  \textbf{2019}(11), 113108 (2019),
\newblock \doi{10.1088/1742-5468/ab4e8e}.

\bibitem{krapivsky-2020}
P.~L. Krapivsky, K.~Mallick and D.~Sels,
\newblock \emph{{Free bosons with a localized source}},
\newblock Journal of Statistical Mechanics: Theory and Experiment
  \textbf{2020}(6), 63101 (2020),
\newblock \doi{10.1088/1742-5468/ab8118}.

\bibitem{rosso-2020}
L.~Rosso, F.~Iemini, M.~Schir{\`o} and L.~Mazza,
\newblock \emph{{Dissipative flow equations}},
\newblock SciPost Phys. \textbf{9}, 91 (2020),
\newblock \doi{10.21468/SciPostPhys.9.6.091}.

\bibitem{vernier-2020}
E.~Vernier,
\newblock \emph{{Mixing times and cutoffs in open quadratic fermionic
  systems}},
\newblock SciPost Phys. \textbf{9}, 49 (2020),
\newblock \doi{10.21468/SciPostPhys.9.4.049}.

\bibitem{alba2021noninteracting}
V.~Alba and F.~Carollo,
\newblock \emph{Noninteracting fermionic systems with localized dissipation:
  Exact results in the hydrodynamic limit},
\newblock arXiv preprint arXiv:2103.05671  (2021).

\bibitem{chaudhari2021zeno}
A.~P. Chaudhari, S.~P. Kelly, R.~J.~V. Tortora and J.~Marino,
\newblock \emph{Zeno crossovers in the entanglement speed of spin chains with
  noisy impurities} (2021), \eprint{2103.16172}.

\bibitem{muller-2021}
T.~{M{\"u}ller}, M.~{Gievers}, H.~{Fr{\"o}ml}, S.~{Diehl} and A.~{Chiocchetta},
\newblock \emph{{Shape effects of localized losses in quantum wires:
  dissipative resonances and nonequilibrium universality}},
\newblock arXiv e-prints arXiv:2105.01059 (2021),
\newblock \eprint{2105.01059}.

\bibitem{gericke-2008}
T.~Gericke, P.~W{\"u}rtz, D.~Reitz, T.~Langen and H.~Ott,
\newblock \emph{High-resolution scanning electron microscopy of an ultracold
  quantum gas},
\newblock Nature Physics \textbf{4}(12), 949 (2008),
\newblock \doi{10.1038/nphys1102}.

\bibitem{brazhnyi-2009}
V.~A. Brazhnyi, V.~V. Konotop, V.~M. P\'erez-Garc\'{\i}a and H.~Ott,
\newblock \emph{Dissipation-induced coherent structures in bose-einstein
  condensates},
\newblock Phys. Rev. Lett. \textbf{102}, 144101 (2009),
\newblock \doi{10.1103/PhysRevLett.102.144101}.

\bibitem{zezyulin-2012}
D.~A. Zezyulin, V.~V. Konotop, G.~Barontini and H.~Ott,
\newblock \emph{Macroscopic zeno effect and stationary flows in nonlinear
  waveguides with localized dissipation},
\newblock Phys. Rev. Lett. \textbf{109}, 020405 (2012),
\newblock \doi{10.1103/PhysRevLett.109.020405}.

\bibitem{barontini-2013}
G.~Barontini, R.~Labouvie, F.~Stubenrauch, A.~Vogler, V.~Guarrera and H.~Ott,
\newblock \emph{Controlling the dynamics of an open many-body quantum system
  with localized dissipation},
\newblock Phys. Rev. Lett. \textbf{110}, 035302 (2013),
\newblock \doi{10.1103/PhysRevLett.110.035302}.

\bibitem{patil-2015}
Y.~S. Patil, S.~Chakram and M.~Vengalattore,
\newblock \emph{Measurement-induced localization of an ultracold lattice gas},
\newblock Phys. Rev. Lett. \textbf{115}, 140402 (2015),
\newblock \doi{10.1103/PhysRevLett.115.140402}.

\bibitem{labouvie-2016}
R.~Labouvie, B.~Santra, S.~Heun and H.~Ott,
\newblock \emph{Bistability in a driven-dissipative superfluid},
\newblock Phys. Rev. Lett. \textbf{116}, 235302 (2016),
\newblock \doi{10.1103/PhysRevLett.116.235302}.

\bibitem{amico2008entanglement}
L.~Amico, R.~Fazio, A.~Osterloh and V.~Vedral,
\newblock \emph{Entanglement in many-body systems},
\newblock Rev. Mod. Phys. \textbf{80}, 517 (2008),
\newblock \doi{10.1103/RevModPhys.80.517}.

\bibitem{calabrese2009entanglemententropy}
P.~Calabrese, J.~Cardy and B.~Doyon,
\newblock \emph{{Entanglement entropy in extended quantum systems}},
\newblock J. Phys. A \textbf{42}(50), 500301 (2009),
\newblock \doi{10.1088/1751-8121/42/50/500301}.

\bibitem{eisert2010colloquium}
J.~Eisert, M.~Cramer and M.~B. Plenio,
\newblock \emph{Colloquium: Area laws for the entanglement entropy},
\newblock Rev. Mod. Phys. \textbf{82}, 277 (2010),
\newblock \doi{10.1103/RevModPhys.82.277}.

\bibitem{laflorencie2016quantum}
N.~Laflorencie,
\newblock \emph{{Quantum entanglement in condensed matter systems}},
\newblock Physics Rep. \textbf{646}, 1 (2016),
\newblock \doi{https://doi.org/10.1016/j.physrep.2016.06.008}.

\bibitem{eisert1999a}
J.~Eisert and M.~B. Plenio,
\newblock \emph{{A comparison of entanglement measures}},
\newblock Journal of Modern Optics \textbf{46}(1), 145 (1999),
\newblock \doi{10.1080/09500349908231260}.

\bibitem{lee2000partial}
J.~Lee, M.~S. Kim, Y.~J. Park and S.~Lee,
\newblock \emph{{Partial teleportation of entanglement in a noisy
  environment}},
\newblock Journal of Modern Optics \textbf{47}(12), 2151 (2000),
\newblock \doi{10.1080/09500340008235138}.

\bibitem{vidal2002computable}
G.~Vidal and R.~F. Werner,
\newblock \emph{Computable measure of entanglement},
\newblock Phys. Rev. A \textbf{65}, 032314 (2002),
\newblock \doi{10.1103/PhysRevA.65.032314}.

\bibitem{plenio2005logarithmic}
M.~B. Plenio,
\newblock \emph{Logarithmic negativity: A full entanglement monotone that is
  not convex},
\newblock Phys. Rev. Lett. \textbf{95}, 090503 (2005),
\newblock \doi{10.1103/PhysRevLett.95.090503}.

\bibitem{ruggiero2016negativity}
P.~Ruggiero, V.~Alba and P.~Calabrese,
\newblock \emph{Negativity spectrum of one-dimensional conformal field
  theories},
\newblock Phys. Rev. B \textbf{94}, 195121 (2016),
\newblock \doi{10.1103/PhysRevB.94.195121}.

\bibitem{ruggiero2016entanglement}
P.~Ruggiero, V.~Alba and P.~Calabrese,
\newblock \emph{Entanglement negativity in random spin chains},
\newblock Phys. Rev. B \textbf{94}, 035152 (2016),
\newblock \doi{10.1103/PhysRevB.94.035152}.

\bibitem{wichterich2009scaling}
H.~Wichterich, J.~Molina-Vilaplana and S.~Bose,
\newblock \emph{Scaling of entanglement between separated blocks in spin chains
  at criticality},
\newblock Phys. Rev. A \textbf{80}, 010304 (2009),
\newblock \doi{10.1103/PhysRevA.80.010304}.

\bibitem{marcovitch2009critical}
S.~Marcovitch, A.~Retzker, M.~B. Plenio and B.~Reznik,
\newblock \emph{Critical and noncritical long-range entanglement in
  klein-gordon fields},
\newblock Phys. Rev. A \textbf{80}, 012325 (2009),
\newblock \doi{10.1103/PhysRevA.80.012325}.

\bibitem{calabrese2012entanglement}
P.~Calabrese, J.~Cardy and E.~Tonni,
\newblock \emph{Entanglement negativity in quantum field theory},
\newblock Phys. Rev. Lett. \textbf{109}, 130502 (2012),
\newblock \doi{10.1103/PhysRevLett.109.130502}.

\bibitem{calabrese2013entanglement}
P.~Calabrese, J.~Cardy and E.~Tonni,
\newblock \emph{Entanglement negativity in extended systems: a field
  theoretical approach},
\newblock Journal of Statistical Mechanics: Theory and Experiment
  \textbf{2013}(02), P02008 (2013).

\bibitem{coser2014entanglement}
A.~Coser, E.~Tonni and P.~Calabrese,
\newblock \emph{Entanglement negativity after a global quantum quench},
\newblock Journal of Statistical Mechanics: Theory and Experiment
  \textbf{2014}(12), P12017 (2014).

\bibitem{eisler2014entanglement}
V.~Eisler and Z.~Zimbor{\'a}s,
\newblock \emph{Entanglement negativity in the harmonic chain out of
  equilibrium},
\newblock New Journal of Physics \textbf{16}(12), 123020 (2014).

\bibitem{eisler2015partial}
V.~Eisler and Z.~Zimbor{\'a}s,
\newblock \emph{On the partial transpose of fermionic gaussian states},
\newblock New Journal of Physics \textbf{17}(5), 053048 (2015).

\bibitem{blondeau2016universal}
O.~Blondeau-Fournier, O.~A. Castro-Alvaredo and B.~Doyon,
\newblock \emph{Universal scaling of the logarithmic negativity in massive
  quantum field theory},
\newblock Journal of Physics A: Mathematical and Theoretical \textbf{49}(12),
  125401 (2016).

\bibitem{shapourian2017many}
H.~Shapourian, K.~Shiozaki and S.~Ryu,
\newblock \emph{Many-body topological invariants for fermionic
  symmetry-protected topological phases},
\newblock Phys. Rev. Lett. \textbf{118}, 216402 (2017),
\newblock \doi{10.1103/PhysRevLett.118.216402}.

\bibitem{shapourian2017partial}
H.~Shapourian, K.~Shiozaki and S.~Ryu,
\newblock \emph{Partial time-reversal transformation and entanglement
  negativity in fermionic systems},
\newblock Phys. Rev. B \textbf{95}, 165101 (2017),
\newblock \doi{10.1103/PhysRevB.95.165101}.

\bibitem{shapourian2019entanglement}
H.~Shapourian and S.~Ryu,
\newblock \emph{Entanglement negativity of fermions: Monotonicity, separability
  criterion, and classification of few-mode states},
\newblock Phys. Rev. A \textbf{99}, 022310 (2019),
\newblock \doi{10.1103/PhysRevA.99.022310}.

\bibitem{eisler2012entanglement}
V.~Eisler and I.~Peschel,
\newblock \emph{On entanglement evolution across defects in critical chains},
\newblock EPL (Europhysics Letters) \textbf{99}(2), 20001 (2012).

\bibitem{collura2013entanglement}
M.~Collura and P.~Calabrese,
\newblock \emph{Entanglement evolution across defects in critical anisotropic
  heisenberg chains},
\newblock Journal of Physics A: Mathematical and Theoretical \textbf{46}(17),
  175001 (2013).

\bibitem{gruber2020time}
M.~Gruber and V.~Eisler,
\newblock \emph{Time evolution of entanglement negativity across a defect},
\newblock Journal of Physics A: Mathematical and Theoretical \textbf{53}(20),
  205301 (2020).

\bibitem{gamayun-2020}
O.~Gamayun, O.~Lychkovskiy and J.-S. Caux,
\newblock \emph{{Fredholm determinants, full counting statistics and Loschmidt
  echo for domain wall profiles in one-dimensional free fermionic chains}},
\newblock SciPost Phys. \textbf{8}, 36 (2020),
\newblock \doi{10.21468/SciPostPhys.8.3.036}.

\bibitem{gamayun2021nonequilibrium}
O.~Gamayun, A.~Slobodeniuk, J.-S. Caux and O.~Lychkovskiy,
\newblock \emph{Nonequilibrium phase transition in transport through a driven
  quantum point contact},
\newblock Phys. Rev. B \textbf{103}, L041405 (2021),
\newblock \doi{10.1103/PhysRevB.103.L041405}.

\bibitem{ptaszy2019entropy}
K.~Ptaszy\'{n}ski and M.~Esposito,
\newblock \emph{Entropy production in open systems: The predominant role of
  intraenvironment correlations},
\newblock Phys. Rev. Lett. \textbf{123}, 200603 (2019),
\newblock \doi{10.1103/PhysRevLett.123.200603}.

\bibitem{burke-2020}
P.~C. Burke, J.~Wiersig and M.~Haque,
\newblock \emph{Non-hermitian scattering on a tight-binding lattice},
\newblock Phys. Rev. A \textbf{102}, 012212 (2020),
\newblock \doi{10.1103/PhysRevA.102.012212}.

\bibitem{degasperis-1974}
A.~Degasperis, L.~Fonda and G.~C. Ghirardi,
\newblock \emph{Does the lifetime of an unstable system depend on the measuring
  apparatus?},
\newblock Il Nuovo Cimento A (1965-1970) \textbf{21}(3), 471 (1974),
\newblock \doi{10.1007/BF02731351}.

\bibitem{misra-1977}
B.~Misra and E.~C.~G. Sudarshan,
\newblock \emph{{The Zeno's paradox in quantum theory}},
\newblock Journal of Mathematical Physics \textbf{18}(4), 756 (1977),
\newblock \doi{10.1063/1.523304}.

\bibitem{facchi-2002}
P.~Facchi and S.~Pascazio,
\newblock \emph{Quantum zeno subspaces},
\newblock Phys. Rev. Lett. \textbf{89}, 080401 (2002),
\newblock \doi{10.1103/PhysRevLett.89.080401}.

\bibitem{peschel2009reduced}
I.~Peschel and V.~Eisler,
\newblock \emph{Reduced density matrices and entanglement entropy in free
  lattice models},
\newblock Journal of physics a: mathematical and theoretical \textbf{42}(50),
  504003 (2009).

\bibitem{audenaert2002entanglement}
K.~Audenaert, J.~Eisert, M.~B. Plenio and R.~F. Werner,
\newblock \emph{Entanglement properties of the harmonic chain},
\newblock Phys. Rev. A \textbf{66}, 042327 (2002),
\newblock \doi{10.1103/PhysRevA.66.042327}.

\bibitem{wong}
R.~Wong,
\newblock \emph{{Asymptotic Approximations of Integrals}},
\newblock Society for Industrial and Applied Mathematics,
\newblock \doi{10.1137/1.9780898719260} (2001).

\bibitem{eisler2011crossover}
V.~Eisler,
\newblock \emph{Crossover between ballistic and diffusive transport: the
  quantum exclusion process},
\newblock Journal of Statistical Mechanics: Theory and Experiment
  \textbf{2011}(06), P06007 (2011).

\bibitem{rossini2021coherent}
D.~Rossini and E.~Vicari,
\newblock \emph{Coherent and dissipative dynamics at quantum phase transitions}
  (2021), \eprint{2103.02626}.

\bibitem{calabrese2012quantum}
P.~Calabrese, F.~H. Essler and M.~Fagotti,
\newblock \emph{Quantum quench in the transverse field ising chain: I. time
  evolution of order parameter correlators},
\newblock Journal of Statistical Mechanics: Theory and Experiment
  \textbf{2012}(07), P07016 (2012).

\end{thebibliography}

\end{document}